\documentclass{emulateapj}
\usepackage{natbib,aas_macros}
\newcommand{\OII}{[O\,{\sc ii}]}
\newcommand{\Ha}{H$\alpha$}
\newcommand{\Hb}{H$\beta$}
\newcommand{\HaNII}{H$\alpha+$[N\,{\sc ii}]}
\newcommand{\CIV}{C{\sc iv}}
\newcommand{\NII}{[N\,{\sc ii}]}
\newcommand{\OIII}{[O\,{\sc iii}]}
\newcommand{\SII}{[S\,{\sc ii}]}
\newcommand{\xmm}{{\it XMM-Newton}}
\newcommand{\fesc}{$f_{\rm esc}^{{\rm Ly}\alpha}$}

\slugcomment{Accepted for publication in ApJ, 2011 Nov 28}

\shorttitle{%
Average Metallicity and SFR of LAEs probed by a Triple NB Survey}
\shortauthors{Nakajima et al.}

\begin{document}

\title{
Average Metallicity and Star Formation Rate of Ly$\alpha$ Emitters 
Probed by a Triple Narrow-Band Survey
\altaffilmark{\dag}
}

\author{Kimihiko Nakajima  \altaffilmark{1,3},
       Masami Ouchi        \altaffilmark{2,3,4,13},
       Kazuhiro Shimasaku  \altaffilmark{1,5},
       Yoshiaki Ono        \altaffilmark{1},
       Janice C. Lee       \altaffilmark{4,13}, \\
       Sebastien Foucaud   \altaffilmark{6}, 
       Chun Ly             \altaffilmark{4,7,8,14},
       Daniel A. Dale      \altaffilmark{9},
       Samir Salim         \altaffilmark{10},
       Rose Finn           \altaffilmark{11}, 
       Omar Almaini        \altaffilmark{12}, \\ and
       Sadanori Okamura    \altaffilmark{1,5}
      }

\email{nakajima@astron.s.u-tokyo.ac.jp}

\altaffiltext{1}{%
Department of Astronomy, Graduate School of Science,
The University of Tokyo, 7-3-1 Hongo, Bunkyo-ku, Tokyo 113-0033,
Japan
}
\altaffiltext{2}{%
Institute for Cosmic Ray Research, The University of Tokyo,
5-1-5 Kashiwanoha, Kashiwa, Chiba 277-8582, Japan
}
\altaffiltext{3}{%
Institute for the Physics and Mathematics of the Universe (IPMU),
TODIAS, The University of Tokyo, 5-1-5 Kashiwanoha, Kashiwa, 
Chiba 277-8583, Japan
}
\altaffiltext{4}{%
Observatories of the Carnegie Institution of Washington,
813 Santa Barbara Street, Pasadena, CA 91101, USA
}
\altaffiltext{5}{%
Research Center for the Early Universe, Graduate School of Science,
The University of Tokyo, Tokyo 113-0033, Japan
}
\altaffiltext{6}{%
Department of Earth Sciences, National Taiwan Normal University, 
N$^{\circ}$88, Tingzhou Road, Sec. 4, Taipei 11677, Taiwan (R.O.C.)
}
\altaffiltext{7}{%
Department of Physics and Astronomy, UCLA, Los Angeles, CA, USA
}
\altaffiltext{8}{%
Space Telescope Science Institute, Baltimore, MD, USA
}
\altaffiltext{9}{%
Department of Physics and Astronomy, University of Wyoming, Laramie,
WY, USA
}
\altaffiltext{10}{%
Department of Astronomy, Indiana University, Bloomington, IN, USA
}
\altaffiltext{11}{%
Department of Physics, Siena College, Loudonville, NY, USA
}
\altaffiltext{12}{%
School of Physics \& Astronomy, University of Nottingham, 
Nottingham, UK
}
\altaffiltext{13}{%
Carnegie Fellow
}
\altaffiltext{14}{%
Giacconi Fellow
}

\altaffiltext{\dag}{%
Based in part on data collected at Subaru Telescope,
which is operated by the National Astronomical Observatory of Japan.
}

%
%

\begin{abstract}

We present the average metallicity and star-formation rate of
Ly$\alpha$ emitters (LAEs)
measured from our large-area survey with three narrow-band (NB) filters
covering the Ly$\alpha$, [O\,{\sc ii}]$\lambda 3727$, and 
H$\alpha$+[N\,{\sc ii}] lines of LAEs at $z=2.2$.
We select 919 $z=2.2$ LAEs from Subaru/Suprime-Cam NB data
in conjunction with Magellan/IMACS spectroscopy. 
Of these LAEs, 561 and 105 are observed with KPNO/NEWFIRM near-infrared
NB filters whose central wavelengths are matched to
redshifted [O\,{\sc ii}] and H$\alpha$ nebular lines, respectively.
By stacking the near-infrared images of the LAEs, 
we successfully obtain average nebular-line fluxes of LAEs,
the majority of which are too faint to be identified 
individually by narrow-band imaging or deep spectroscopy.
The stacked object has an H$\alpha$ luminosity of 
$1.7\times 10^{42}$\,erg\,s$^{-1}$
corresponding to a star formation rate (SFR) of $14\,M_{\odot}$\,yr$^{-1}$.
We place, for the first time, a firm lower limit to 
the average metallicity of LAEs of $Z\gtrsim 0.09\,Z_{\odot}$ ($2\sigma$)
based on the [O\,{\sc ii}]/(H$\alpha$+[N\,{\sc ii}]) index 
together with photo-ionization models and empirical relations.
This lower limit of metallicity rules out
the hypothesis that LAEs, so far observed at $z\sim 2$,
are extremely metal poor ($Z<2\times 10^{-2}\,Z_{\odot}$) galaxies 
at the $4\sigma$ level.
This limit is higher than a simple extrapolation of
the observed mass-metallicity relation of
$z\sim 2$ UV-selected galaxies toward lower masses
($5\times 10^8\,M_{\odot}$), 
but roughly consistent with a recently proposed fundamental
mass-metallicity relation when the LAEs' relatively low SFR
is taken into account. The H$\alpha$ and Ly$\alpha$ luminosities
of our NB-selected LAEs indicate that
the escape fraction of Ly$\alpha$ photons
is $\sim 12-30$\,\%, much higher than the values
derived for other galaxy populations at $z\sim 2$.

\end{abstract}


\section{INTRODUCTION} \label{sec:introduction}

Galaxy mass is thought to be a fundamental quantity which governs
the evolution of galaxies.
Detailed observations of present-day galaxies show that
various properties of galaxies such as star formation rate (SFR) and
gas phase metallicity correlate with mass
(e.g., \citealt{brinchmann2004,tremonti2004,ellison2008}).
Theoretically, galaxy formation models based on $\Lambda$CDM cosmology
predict that galaxies grow through subsequent mergings of lower-mass
objects and that galaxy properties are largely determined
by their masses through mass-dependent processes at work
in the evolution of galaxies
(e.g., \citealt{blumenthal1984,davis1985,bardeen1986}).
Therefore, observations of the mass-dependence of galaxy properties
back in cosmic time are crucial to understanding how galaxies evolve
to acquire present-day properties.

Past observations have revealed that some physical quantities
like SFR and metallicity correlate with mass also at high-$z$, 
although the correlation seems to evolve with redshift 
relative to the local galaxies 
(e.g., \citealt{reddy2006,hayashi2009,erb2006a,maiolino2008,mannucci2009}).
They demonstrated the importance of mass-dependent effects on galaxy evolution.
However, high-$z$ samples are biased for high-mass galaxies
($\gtrsim 10^9\,M_\odot$) because most of them are based on
continuum selected samples,
such as Lyman-break galaxies (LBGs; e.g., \citealt{SH1992}) 
and BzK galaxies (e.g., \citealt{daddi2004}).

\begin{deluxetable*}{lccccl}
\tablecolumns{6}
\tabletypesize{\scriptsize}
\tablecaption{Summary of NB387 Imaging Data in the SXDS field %
\label{tbl:sum_ndata}}
\tablewidth{420pt}
\setlength{\tabcolsep}{0.2in}
\tablehead{%
\colhead{Field Name} &
\colhead{Exp. Time} &
\colhead{PSF} &
\colhead{Area} &
\colhead{$m_{\rm lim}$} &
\colhead{Date} \\
\colhead{} &
\colhead{(1)} &
\colhead{(2)} &
\colhead{(3)} &
\colhead{(4)} &
\colhead{}
}
\startdata
SXDS-C & 3.20 [9] & 0.88 & 587 [41] & 25.7 [25.3] & 2009 Dec 14 - 16 \\
SXDS-N & 2.50 [5] & 0.70 & 409 [159] & 25.6 [25.2] & 2009 Dec 16 \\
SXDS-S & 2.50 [5] & 0.85 & 775 [344] & 25.7 [25.3] & 2009 Dec 16 \\
SXDS-E\tablenotemark{(5)} & 3.33 [10] & 1.95 & \nodata & \nodata & 2009 Dec 19, 20 \\
SXDS-W & 1.83 [5] & 1.23 & 232 [122] & 25.1 [24.7] & 2009 Dec 16, 19
\enddata
\tablenotetext{(1)}{%
Total exposure times (hour).
The value in square brackets shows the numbers of exposures 
that are combined.
}
\tablenotetext{(2)}{%
FWHM of PSFs of the stacked image that are registered with broadband images 
(arcsec).
}
\tablenotetext{(3)}{%
Effective area that is used for the selection (arcmin$^2$).
The value in square brackets shows the area with low-$S/N$.
}
\tablenotetext{(4)}{%
Limiting magnitude derived from $5\sigma$ sky noise in a $2\arcsec$
 diameter aperture.
Note that $m_{\rm lim}$ in the SXDS-W is defined in a $2\farcs 5$
 diameter aperture.
The magnitude in square brackets shows the limiting magnitudes in the
 low-$S/N$ regions.
}
\tablenotetext{(5)}{%
The SXDS-E has large PSF size and is not included in the following analysis.
So, we do not report the $m_{\rm lim}$ and covered area in this field.
}
\end{deluxetable*}

Ly$\alpha$ emitters (LAEs), galaxies commonly observed at high redshifts
with strong Ly$\alpha$ emission,
are likely to be low-mass, young galaxies
as suggested from their small sizes, faint continua,
and low stellar masses inferred from spectral energy
distribution (SED) fitting
($\lesssim 10^9\,M_\odot$; \citealt{gawiser2006,gawiser2007,%
finkelstein2007,finkelstein2008,finkelstein2009,%
nilsson2007,pirzkal2007,lai2008,ono2010a,ono2010b,yuma2010}).
Since they can be efficiently detected by narrow-band imaging,
LAEs are a useful probe to investigate low-mass galaxies
in the early stages of galaxy evolution.
Furthermore, low-mass galaxies at high redshifts such as LAEs 
are especially interesting since they are likely to be building 
blocks of massive galaxies seen in later epochs.

Thanks to the remarkable progress in observations of LAEs,
our knowledge of their properties is rapidly accumulating
(e.g., \citealt{CH1998,MR2002,ouchi2003,MR2004,gawiser2006,%
kashikawa2006,shimasaku2006,gronwall2007,ouchi2008,%
nilsson2009,blanc2011,guaita2010,hayes2010,ono2010a,ono2010b,%
ouchi2010,finkelstein2011,guaita2011,nilsson2011}).
However, in almost all the observations, physical properties of LAEs
including SFR and metallicity have been generally estimated 
by the SED fitting of broadband photometry.
This is in contrast to massive galaxies at similar redshifts. 
While a large number of massive, continuum-selected galaxies 
have now direct measurements of SFR and metallicity from
nebular lines (e.g., \citealt{reddy2006,erb2006b,maiolino2008,%
mannucci2009,hayashi2009,yoshikawa2010}), 
there are only a few LAEs with such direct measurements 
\citep{mclinden2011,finkelstein2011}.
One of the major reasons for the paucity of the direct measurements 
is that well-studied LAEs are located at very high redshift ($3<z<7$), 
where (rest-frame optical) nebular lines redshift into infrared wavelengths 
that cannot be observed from the ground. 
However, nebular lines of bright LAEs at moderate redshifts ($z=2-3$)
have now been measured through recent LAE surveys 
(e.g., \citealt{mclinden2011,finkelstein2011}).
It is known that SFRs derived from SED fitting are dependent on
the star formation history assumed, and they can vary by an order of
magnitude among different histories (e.g., \citealt{ono2010a}).
It is also known that SED fitting cannot strongly constrain
metallicities due to the degeneracy with age (e.g., \citealt{ono2010a}).
Observations of nebular lines for a large number of LAEs
are essential to extending star formation rate and metallicity
measurements toward low-mass galaxies below
$\sim 10^9\,M_\odot$, so that the mass-dependencies
of SFR and metallicity can be compared with their present-day homologue 
over a full mass range.

We are conducting an imaging survey of $z\simeq 2.2$ LAEs
in several fields on the sky using three narrow-band filters described below.
This redshift is unique because \OII$\lambda3727$ and \Ha\
lines fall into wavelength ranges where OH-airglow is very weak,
thus enabling one to study SFRs and metallicities of LAEs using these
lines from the ground.
We developed a new narrow-band filter, NB387, with a central wavelength
and FWHM of $3870$\,\AA\ and $94$\,\AA, respectively, to select
LAEs over $z=2.14$ -- $2.22$.
\OII\ and \Ha\ lines in this redshift range are then observed
through near-infrared (NIR) narrow-band filters, 
NB118 ($\lambda_c=11866$\,\AA, FWHM$=111$\,\AA) and
NB209 ($\lambda_c=20958$\,\AA, FWHM$=205$\,\AA), respectively,
developed by the NewH$\alpha$ Survey (Lee et al. in preparation).
In this paper, we present the results from
data of the Subaru/\xmm\ Deep Survey field,
which are the first results of this triple-narrowband survey.

While our NB118 and NB209 imaging in the SXDS is not deep enough
to detect \OII\ and \Ha\ lines for individual objects except for
very luminous ones, we successfully detect these emission lines
in the stacked images of more than $100$ LAEs.
We then measure their fluxes to derive the SFR and metallicity,
and place the constraints on the average SFR and metallicity 
of a typical LAE at $z=2.2$ for the first time. 
We use these measurements to discuss the SFR and metallicity of
low-mass ($<10^9\,M_\odot$) galaxies at $z\sim 2$.

\begin{figure*}
\epsscale{1.05}
\plotone{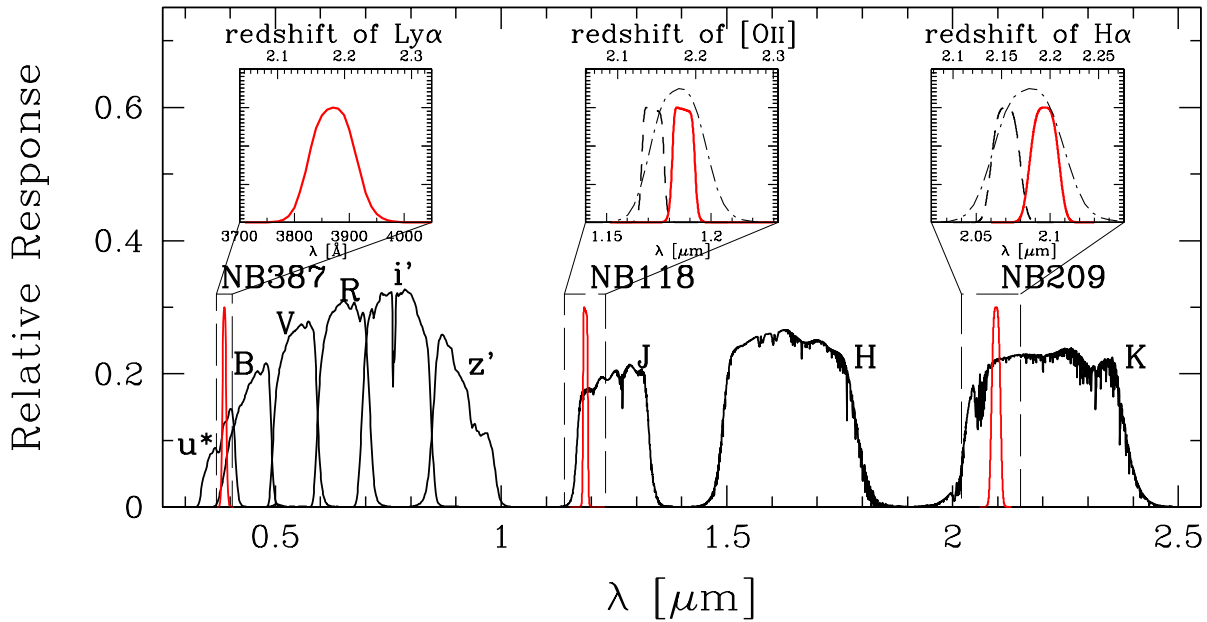}
\caption{
Relative response curves of the triple narrow-band filters, 
NB387 at Subaru and NB118 and NB209 at KPNO/NEWFIRM (red), 
superimposed on those of 
CFHT $u^*$, Subaru $B,V,R,i',z'$, and UKIRT/WFCAM $J, H, K$ (black).
The responses include the throughputs of the instrument and
the telescope as well as atmospheric absorption.
The inner panels are a zoom in around NB387, NB118, and NB209, 
whose upper $x$-axis shows the redshift of Ly$\alpha$, \OII, and 
\Ha, respectively. 
In the NB118 and NB209 zoom-in panels, 
the dashed curves indicate the response curves of NB118 and NB209 
at the corner of the filter 
(see text for details; Equation (\ref{eq:NIR_NB_shift})), 
while the dot-dashed lines are re-scaled response curves of 
NB387 along the wavelength axis to sample \OII\ and \Ha\ 
after correction for 
a velocity offset between Ly$\alpha$ and nebular lines of
$400$\,km\,s$^{-1}$.
\label{fig:filters_all}}
\end{figure*}

Since Ly$\alpha$ photons produced in a galaxy are expected
to be easily absorbed by dust in the interstellar medium (ISM) of the galaxy
during repeated resonant scatterings by neutral hydrogen gas,
a puzzle is why LAEs have such strong Ly$\alpha$ emission.
There are three possible answers to the puzzle. 
The first is that LAEs are primordial galaxies 
(e.g., \citealt{scannapieco2003}). If population III stars 
are formed following a top-heavy initial mass function, 
hard UV photons would be radiated in a short time scale, 
and strong Ly$\alpha$ lines would be observed. 
In this case, LAEs are young, and extremely metal poor galaxies.
This will be tested by the metallicity we estimate.
The second is that the ISM has a clumpy geometry
(e.g., \citealt{neufeld1991}), where the Ly$\alpha$ photons are scattered
at the surfaces of the clumps and thus are not heavily absorbed before
escaping from the galaxy. 
The third is the outflow of the ISM (e.g., \citealt{kunth1998}). 
Ly$\alpha$ photons that are scattered at the far side of expanding ISM 
can be Doppler shifted to have redder wavelengths, and escape from 
galaxies without being heavily absorbed by neutral hydrogen gas. 
In order to investigate these possibilities, 
we estimate the Ly$\alpha$ escape fraction, \fesc, for our LAEs
by comparing the observed Ly$\alpha$ luminosity with the intrinsic
Ly$\alpha$ luminosity predicted from the dust-corrected \Ha\ luminosity.

This paper is organized as follows.
We describe the data in \S \ref{sec:data}.
The sample of $z=2.2$ LAEs is constructed in \S \ref{sec:selection}, 
where the results of optical spectroscopy are also shown.
In \S \ref{sec:detections}, we detect \OII\ and \Ha\ emission
in the stacked LAEs and calculate their equivalent widths.
Objects with individual detections of these lines are also briefly
mentioned.
Results of SED fitting of the stacked LAEs are briefly described 
in \S \ref{sec:SED_fitting}.
In \S \ref{sec:discussion},
we derive SFR, metallicity, and \fesc\ for our LAEs,
and discuss their implications.
Conclusions are given in \S \ref{sec:conclusion}.
Throughout this paper, magnitudes are given in the AB system
\citep{oke1974},
and we assume a standard $\Lambda$CDM cosmology with
$(\Omega_m,\Omega_{\Lambda},H_0)=
(0.3,0.7,70\,{\rm km}\,{\rm s}^{-1}\,{\rm Mpc}^{-1})$

\section{IMAGING DATA} \label{sec:data}

In this section we describe the optical and NIR data
in the SXDS field used in our analysis.

\begin{deluxetable*}{lccccc}
\tablecolumns{6}
\tabletypesize{\scriptsize}
\tablecaption{Summary of Optical Broadband Imaging Data in the SXDS field %
\label{tbl:sum_bdata}}
\tablewidth{400pt}
\setlength{\tabcolsep}{0.17in}
\tablehead{%
\colhead{Band} &
\colhead{Observatory} &
\colhead{Field Name} &
\colhead{PSF} &
\colhead{$m_{\rm lim}$} & 
\colhead{Reference} \\
\colhead{} &
\colhead{} &
\colhead{} &
\colhead{(1)} &
\colhead{(2)} &
\colhead{(3)}
}
\startdata
$u^*$\tablenotemark{(4)} & CFHT   & SXDS-C,N,S,E,W & $0.85$ & $26.9$ & (a) \\
$B$     & Subaru & SXDS-C,N,S,E,W & $0.78 - 0.84$ & $27.5 - 27.8$ & (b) \\
$V$     & Subaru & SXDS-C,N,S,E,W & $0.72 - 0.82$ & $27.1 - 27.2$ & (b) \\
$R$     & Subaru & SXDS-C,N,S,E,W & $0.74 - 0.82$ & $27.0 - 27.2$ & (b) \\
$i'$    & Subaru & SXDS-C,N,S,E,W & $0.68 - 0.82$ & $26.9 - 27.1$ & (b) \\
$z'$    & Subaru & SXDS-C,N,S,E,W & $0.70 - 0.76$ & $25.8 - 26.1$ & (b)
\enddata
\tablenotetext{(1)}{%
PSF size is defined as a FWHM of point sources (arcsec).
}
\tablenotetext{(2)}{%
The limiting magnitude ($5\sigma$) estimated by $2\arcsec$ diameter
random aperture photometry.
}
\tablenotetext{(3)}{%
(a) Foucaud et al. in preparation, (b) \citet{furusawa2008}
}
\tablenotetext{(4)}{%
The $u^*$ image covers the same area of the UKIDSS/UDS project
\citep{lawrence2007} which corresponds to about $77$\,\% of the
SXDS-C,-N,-S,-E, and -W.
}
\end{deluxetable*}

\subsection{NB387 Images} \label{ssec:OPT_NB}

We carried out NB387 imaging observations of the SXDS
with Subaru/Suprime-Cam \citep{miyazaki2002}
on 2009 December 14-16 and 19-20.
Table \ref{tbl:sum_ndata} summarizes the details of the observations.
The SXDS field is covered by deep Suprime-Cam broadband data in five
pointings with small overlaps; these five {\lq}sub-fields{\rq} are
named SXDS-C, -N, -S, -E, and -W, respectively,
after their relative positions on the sky \citep{furusawa2008}.
We acquired NB387 imaging for all the five sub-fields
(Table \ref{tbl:sum_ndata}).
We do not, however, use the data of SXDS-E in this study
because of bad seeing.
For photometric calibration,
we observed spectrophotometric standard stars Feige34,
LDS749B, and G93-48 \citep{oke1990}.
Each standard star was observed more than twice
under photometric condition with airmasses of $1.1 - 1.3$.

We used the Suprime-Cam Deep Field Reduction package
\citep[SDFRED;][]{yagi2002,ouchi2004}
to reduce the NB387 data.
The data reduction process included bias subtraction,
flat fielding,
distortion correction,
cosmic ray rejection,
sky subtraction,
bad pixel/satellite trail masking,
image shifting, and stacking.
For cosmic ray rejection, we used {\it LA.COSMIC} \citep{vandokkum2001}.
After the stacking process, our images were registered 
with the archival broadband images (\S \ref{ssec:OPT_BB})
using bright stellar objects commonly detected in the NB387 and
the broadband images.

The PSF sizes of the registered images for the four sub-fields
are $0\farcs 70 - 1\farcs 2$.
The $5 \sigma$ detection limits in a $2\arcsec$ diameter aperture
are $25.1-25.7$ mag except on the edges of the images
where signal-to-noise ratios ($S/N$) were significantly lower
due to dithering.
However, we include those low $S/N$ regions to increase the number
of LAEs, setting brighter limiting magnitudes according to
the $S/N$ ratios.
The limiting magnitudes are estimated in the same manner as
in \citet{furusawa2008}; we spread $5,000$ $2\farcs 0$ diameter apertures
over the entire image randomly after masking detected objects, 
and measure their photon counts. 
We then fit the negative part of the histogram of 
the counts with Gaussian, and regard its $\sigma$ as
the $1\sigma$ sky fluctuations of the image for $2\farcs 0$ 
diameter apertures. The limiting magnitudes which appear in the following 
subsections are also estimated in the same manner.

We infer the errors in photometric zero points of
our NB387 images from a comparison of colors ($u^*-$NB387 or $B-$NB387) 
of stellar objects in the images with those of $175$ Galactic
stars calculated from \cite{GS1983}'s spectrophotometric atlas.
The estimated errors are less than $0.05$ mag, which are small
enough for our study. 
\begin{figure}
\epsscale{1.15}
\plotone{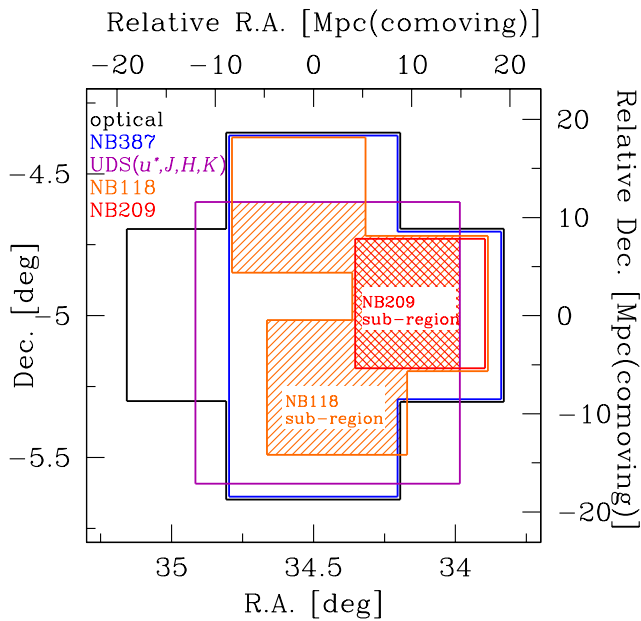}
\caption{
Areas covered by respective imaging data in the SXDS.
Lines in different colors outline areas covered in different
passbands: blue: NB387; magenta: $u^*$ and $J,H,K$;
orange: NB118; red: NB209; black: optical ($B,V,R,i',z'$).
The orange-shaded and red-shaded regions correspond to
the NB118 sub-region and the NB209 sub-region, respectively.
Note that the NB209 sub-region is embedded in the the NB118 sub-region
(see \S \ref{sec:selection}).
\label{fig:fov_sxds}}
\end{figure}

\subsection{Optical Broadband Images} \label{ssec:OPT_BB}

The optical broadband data are required not only for selecting LAEs
but also for performing SED fitting of the stacked object 
(Y. Ono et al. in preparation; briefly mentioned in \S \ref{sec:SED_fitting}).
We use the publicly available $B$, $V$, $R$, $i'$, and $z'$ data
taken with Subaru/Suprime-Cam by 
the SXDS project \citep{furusawa2008},
and the $u^*$ data taken with CFHT/MegaCam as part of
the UKIDSS/UDS project (Foucaud et al. in preparation).
Table \ref{tbl:sum_bdata} summarizes the details of 
the optical broadband data. 
The passbands of $u^*$ and $B$, which are used to select LAEs
as off-bands of NB387, are shown in Figure \ref{fig:filters_all}.
We register the $u^*$ image with the Suprime-Cam $B$
image using common bright stars.
The $5\sigma$ limiting magnitudes on a $2\arcsec$ diameter aperture 
are: 
$26.9\,(u^*)$, $27.5 - 27.8\,(B)$, 
$27.1 - 27.2\,(V)$, $27.0 - 27.2\,(R)$, 
$26.9 - 27.1\,(i')$, $25.8 - 26.1\,(z')$ mag. 
The $u^*$ image covers $77$\,\% of the Suprime-Cam field,
nearly the same area as that of the UKIDSS/UDS $JHK$ images
(\citealt{lawrence2007}; see \S \ref{ssec:NIR_BB})
Figure \ref{fig:fov_sxds} illustrates
the sky coverages of our NB387 imaging,
the Suprime-Cam broadband imaging, and the $u^*$ imaging.

For each sub-field, PSF sizes were matched to the worst one
among the $u^*BVRi'z'$ and NB387 images
with the IRAF task \verb+GAUSS+.
The resulting PSF FWHMs of the images in the SXDS-C,N,S,W are
$0\farcs 88$ (SXDS-C), $0\farcs 85$ (N), $0\farcs 85$ (S), 
and $1\farcs 23$ (W).

\subsection{NB118 and NB209 Images} \label{ssec:NIR_NB}

The SXDS field has been partly imaged in the NIR NB118 and NB209 
narrowbands with KPNO/NEWFIRM by the NewH$\alpha$ Survey 
(Lee et al. in preparation). 
More details on the NB118 observations are given in \citet{ly2011}.
The regions imaged are shown in Figure \ref{fig:fov_sxds}.
The response curves of these two narrowband filters are shown
in Figure \ref{fig:filters_all} with those of
the WFCAM $J$, $H$, and $K$ bands (\S \ref{ssec:NIR_BB}).
Exposure times were $8.47 - 12.67$ hr for NB118
and $11.75$ hr for NB209.
Both images were registered to the Suprime-Cam 
$z'$ band images using common bright stars%
\footnote{%
in order to run SExtractor in double-image mode 
(see \S \ref{ssec:candidate}).%
}.
The $5\sigma$ limiting magnitudes
in a $2\arcsec$ diameter aperture 
are estimated to be $\simeq 23.6$ mag for the NB118 
and $\simeq 22.6$ mag for the NB209 \footnote{%
Our effective limiting magnitudes of NB118 and NB209 
are $0.2-0.3$ mag deeper than the magnitudes listed here 
owing to smoothing procedures (see \S \ref{ssec:detect_stacking}).
The limiting magnitudes after the smoothing are described 
in \S \ref{ssec:detect_individual}.
} .
Errors in photometric zero points of the NB118 and NB209 images 
are less than $0.05$ mag, which are inferred following 
the same manner as the NB387 images.

In NEWFIRM the incident angle to the filter surface 
is not exactly normal but varies
as a function of the distance from the field center
(Lee et al. in preparation; 
see also \citet{tanaka2011} for Subaru/MOIRCS).
Accordingly, the central wavelength of NB118 and NB209 also
varies over the FoV, since both filters are interference filters.
The central wavelength at an incident angle $\theta$
(angle from normal incidence) is given by
\begin{eqnarray}
\lambda(\theta) = \lambda_0\cos(\theta/n) \label{eq:NIR_NB_shift},
\end{eqnarray}
where $\lambda_0$ is the wavelength at normal incidence
and $n$ is the index of refraction of the material 
on which interference film is coated.
We adopt $n=1.50$ for NB118 and $n=1.49$ for NB209 
(Lee et al. in preparation).
At the corner of the filter $\theta$ has the maximum value
of $13.5$ degree.
Thus, the maximum wavelength shifts are
$\simeq -150$\,\AA\ and $\simeq -260$\,\AA\ for NB118 and NB209,
respectively.
The passband of NB387 is broad enough to
cover Ly$\alpha$ lines at the redshifts of \OII\ and \Ha\
lines corresponding to these shifted wavelengths 
(see also inner panels of Figure \ref{fig:filters_all}).
However, as seen in \S \ref{sssec:MCS},
we have to take into account the passband shift over the FoV
when calculating the equivalent widths of
\OII\ and \Ha\ lines of stacked objects.

\begin{figure}
\epsscale{1.15}
\plotone{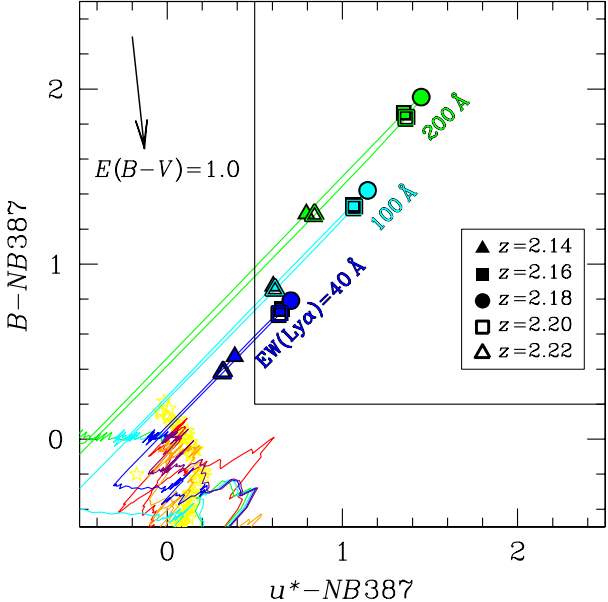}
\caption{
$B-$NB$387$ vs. $u^*-$NB$387$ plane for selection of LAEs at $z \simeq 2.2$.
The solid lines in various colors indicate tracks of model galaxies
redshifted from $0$ to $3.5$
with and without Ly$\alpha$ emission:
purple: simple stellar population with an age of $0.03$\,Gyr;
green, cyan, and blue: simple stellar population with Ly$\alpha$
emission of EW$_{\rm rest}=200, 100$, and $40$\,\AA, respectively.
The symbols on the tracks of model galaxies with Ly$\alpha$ emission 
correspond to $z=2.14$ (filled triangles), $2.16$ (filled squares),
$2.18$ (filled circles), $2.20$ (open squares), and $2.22$ (open triangles).
The red and orange solid lines represent tracks of 
elliptical and spiral galaxies from the SWIRE template library 
\citep{polletta2007}, respectively.
The yellow stars show Galactic stars from \citet{GS1983}.
The tilted arrow indicates the reddening effect in the case of
$E(B-V)=1.0$ \citep{calzetti2000}.
The objects that are located in the area enclosed by the solid black lines
are considered to be LAE candidates.
\label{fig:cc_diagram_model}}
\end{figure}

\subsection{NIR Broadband Images} \label{ssec:NIR_BB}

The UKIDSS/UDS project provides deep $J$, $H$, and $K$
images of UKIRT/WFCAM \citep{lawrence2007}.
In this paper, we use the data release 8 (DR8) images
currently available to the UKIDSS consortium.
The $J$ and $K$ images are used as off-bands of NB118 and NB209
to detect \OII\ and \Ha($+$\NII) emission, respectively,
while the $H$ image is used for SED fitting of the stacked objects.
We register these three images with the Suprime-Cam $z'$ band images
in the same manner as for NB118 and NB209.
The $5\sigma$ limiting magnitudes over a $2\arcsec$ diameter aperture
are estimated to be $24.8$, $24.1$, and $24.6$ in the $J$, $H$, and
$K$ bands, respectively.
$J$, $H$, and $K$ response curves
are shown in Figure \ref{fig:filters_all}.

\section{PHOTOMETRIC SAMPLES OF LAEs AT $z=2.2$} \label{sec:selection}

We select LAEs in a $2,003$ arcmin$^2$ region which is covered
by all the three passbands for selecting LAEs:
NB387, $u^*$, and $B$ (see Figure \ref{fig:fov_sxds}).
In this paper, however, we use only LAEs in a sub-region of
$1,283$ arcmin$^2$ which is also covered by NB118.
In this {\lq}NB118 sub-region{\rq}, a $353$ arcmin$^2$ region is
covered by NB209 as well ({\lq}NB209 sub-region{\rq}).
Note that the $J,H,K$ data are available for the NB118 sub-region.
We use LAEs in the NB118 sub-region to derive the typical \OII\
luminosity of LAEs, and those in the NB209 sub-region to derive
the typical \Ha\ luminosity of LAEs.
The typical \OII\ luminosity is also derived for LAEs in the
NB209 sub-region to estimate the metallicity in combination
with the \Ha\ luminosity.
In this section, we describe the construction of the LAE sample
for the entire $2,003$ arcmin$^2$ region.

\subsection{Object Detection and Candidate Selection} \label{ssec:candidate}

We use the SExtractor software \citep{BA1996} to perform source
detection and photometry.
We run SExtractor in double-image mode, with the NB387 image 
used as the detection image.
We identify sources with $5$ adjoining pixels and 
brightness above $>2\sigma$ of the sky background.
The diameter to define aperture magnitudes is set to
$2\farcs 5$ for the SXDS-W and $2\farcs 0$ for the other sub-fields.
We use aperture magnitudes to calculate colors, and adopt 
\verb+MAG_AUTO+ for the total magnitude.
All magnitudes are corrected for Galactic extinction of $E(B-V)=0.020$
\citep{schlegel1998}.
The NB387-detection catalog contains $42,995$ objects
with aperture magnitudes brighter than the $5\sigma$ sky noise.

We select LAE candidates on the $u^*-$NB387 vs. $B-$NB387 color plane
(Figure \ref{fig:cc_diagram_model}).
In this figure, colors of model galaxies and Galactic stars
are plotted in order to define the selection criteria for LAEs.
The tracks indicate the colors of
model galaxies redshifted from $0.00$ to $3.50$
with a step of $\Delta z=0.01$.
We assume the intergalactic medium (IGM) attenuation model 
of \citet{madau1995}.
Base on Figures \ref{fig:cc_diagram_model} and \ref{fig:cm_diagram_data}, 
we define the color criteria of $z\sim 2.2$ LAEs as:
\begin{eqnarray}
u^*-{\rm NB387} > 0.5 \ \&\&\ B-{\rm NB387} > 0.2, \label{eq:LAE_thresh}
\end{eqnarray}
which select LAEs with EW$_{\rm rest} \gtrsim 30$\,\AA.
The $2\sigma$ photometric errors in $u^*-$NB387 for the faintest 
objects (NB387=$25.7$) in our NB387-detected catalog are $\simeq 0.5$ mag.
Thus, the criterion of $u^*-$NB387$>0.5$ ensures that the contamination 
fraction in our LAE sample due to photometric errors is 
sufficiently low.
We use $2\sigma$ limiting magnitudes instead 
when an object is not detected in $u^*$ or $B$ at $2\sigma$ level. 
The selection criteria require NB387 magnitude significantly 
brighter than both $u^*$ and $B$ magnitudes, which results in 
small number of non-emitters in the sample.

Using Equation (\ref{eq:LAE_thresh}), we identify $1,044$ LAE candidates
in the NB387 detected catalog.
These candidates are contaminated by spurious objects
and foreground and background interlopers.
We remove those contaminants by the procedures described
in the next two subsections.

\subsection{Rejection of Spurious Objects} \label{ssec:spurious}

Given the limited number of ditherings in our NB387 imaging ($\sim 4-5$), 
a clipped-mean stacking fails to completely remove the remaining 
cosmic rays which survived {\it LA.COSMIC} rejection on 
individual dithered images.
Since cosmic rays have steeper light profiles than the PSF, 
we removed $21$ sources with FWHMs significantly smaller than the PSF.
We then perform visual inspection on all the remaining objects,
and eliminate $90$ obvious spurious sources such as ghosts due to
bright stars and bad pixels.

\begin{figure}
\epsscale{1.15}
\plotone{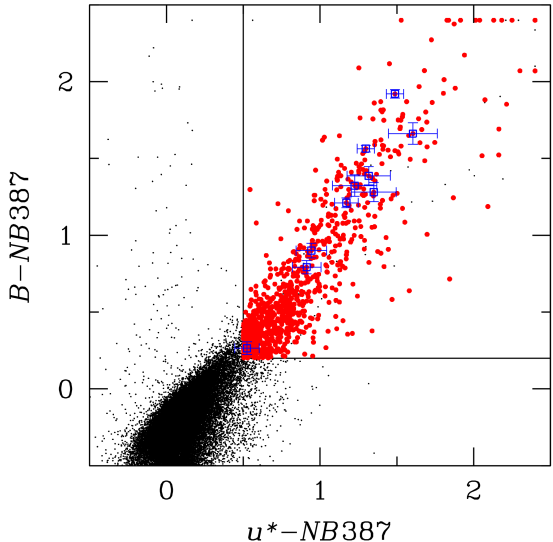}
\caption{
Distribution of NB387-detected objects
in the $B-$NB$387$ vs. $u^*-$NB$387$ plane.
The black dots indicate all the detected objects, while
the red filled circles show LAE candidates
after removing spurious objects and interlopers.
The blue open squares with errorbars show 
spectroscopically confirmed LAEs.
For the purpose of display, objects whose $u^*-$NB$387$ colors 
exceed $2.4$ are plotted at $u^*-$NB$387=2.4$.
\label{fig:cc_diagram_data}}
\end{figure}

\subsection{Identification of Interlopers} \label{ssec:interloper}

In addition to LAEs, other emission line objects, such as
\OII\ emitters at $z\simeq 0.04$, Mg\,{\sc ii}$\lambda 2798$ emitters 
at $z\simeq 0.4$, and \CIV$\lambda 1550$ emitters at $z\simeq 1.5$,
may be included in our sample.
Our survey area, $2,003\,{\rm arcmin}^2$, corresponds to
$400\,{\rm Mpc}^3$ for \OII\ emitters,
which is two orders of magnitude smaller than the volume sampled 
by Ly$\alpha$ emitters ($48,100\, {\rm Mpc}^3$).
The number of \OII\ emitters is therefore expected to be small.

To remove \OII\ emitters from our sample,
we use the Galaxy Evolution Explorer (GALEX)
NUV ($\lambda_c = 2267$\,\AA , FWHM = $616$\,\AA) and
FUV ($\lambda_c = 1516$\,\AA , FWHM = $269$\,\AA) data.
Real LAEs in our sample will be invisible in these data,
since these two passbands are located shortward of
the Lyman break at $z\sim 2.2$.
Thus, objects visible in either of the two GALEX band data
are likely to be \OII\ emitters.
Although some studies have shown that ionizing photons are 
more likely to escape from Ly$\alpha$ selected galaxies than from 
UV-selected galaxies (e.g., \citealt{iwata2009,nestor2011}), 
the estimated UV-to-Ly-continuum flux density ratio 
is $\gtrsim 2$ even for LAEs \citep{nestor2011}, 
therefore the Ly-continuum of $z\sim 2$ LAEs is expected to be 
fainter than the UV continuum by $\sim 1$ mag or more.
Since the detection limit of the GALEX images is $\sim 24$ 
($3\sigma$) both in NUV and FUV, 
LAEs fainter than $V\sim 23$ should be invisible in the GALEX images. 
We find $12$ objects which have a counterpart in either of
the NUV or FUV image within $3\arcsec$ from the NB387 position.
Among them, $4$ are fainter than $V=23$, thus to be inferred 
to be interlopers.
The remaining eight are brighter than $V=23$, and 
their GALEX magnitudes are also bright enough to be 
consistent with them being interlopers.
Indeed, all eight are also detected as 
an X-ray or radio source as described in the next paragraph. 
We thus remove these $12$ objects from the sample.

\begin{figure}
\epsscale{1.15}
\plotone{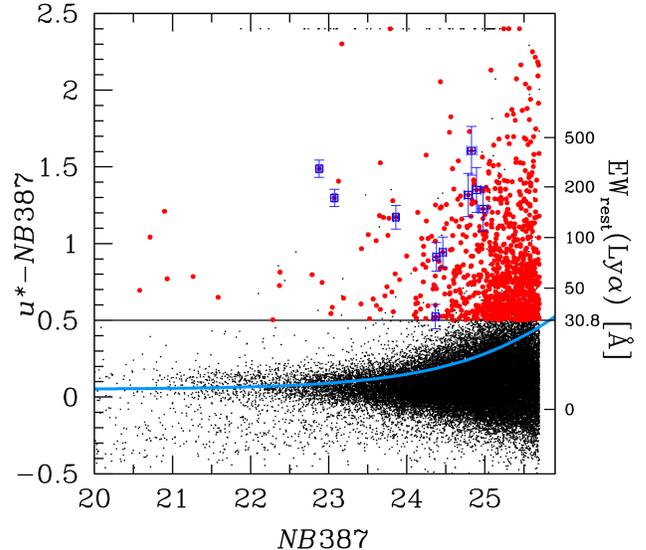}
\caption{
Distribution of NB387-detected objects
in the $u^*-$NB$387$ vs. NB387 plane.
The black dots indicate all the detected objects, while
the red filled circles show LAE candidates.
The blue open squares with errorbars show
spectroscopically confirmed LAEs.
For the purpose of display, objects whose $u^*-$NB$387$ colors
exceed $2.4$ are plotted at $u^*-$NB387$=2.4$.
The horizontal solid line shows the selection threshold of
$u^*-$NB$387$ and the blue curve indicates the $2\sigma$ 
photometric error in $u^*-$NB$387$ for sources with $u^*-$NB$387=0.05$, 
which is the average $u^*-$NB$387$ color of all the objects.
The right $y$ axis shows the rest-frame Ly$\alpha$ equivalent
width of $z=2.18$ LAEs with $u^*-$NB$387$ color
corresponding to the left $y$ axis.
\label{fig:cm_diagram_data}}
\end{figure}

\begin{deluxetable*}{ccccccc}
\tablecolumns{7}
\tabletypesize{\scriptsize}
\tablecaption{Properties of the spectroscopic sample
\label{tbl:Lya_spectra}}
\tablewidth{400pt}
\setlength{\tabcolsep}{0.13in}
\tablehead{
\colhead{ID} &
\colhead{R.A.\tablenotemark{(1)}} &
\colhead{Dec.\tablenotemark{(1)}} &
\colhead{mag(NB387)} &
\colhead{$\lambda_{\rm obs}$} &
\colhead{$z$} &
\colhead{flag} \\
\colhead{} &
\colhead{} &
\colhead{} &
\colhead{(2)} &
\colhead{(3)} &
\colhead{(4)} &
\colhead{(5)}  
}
\startdata
NB387-C-04640 & 02:18:48.968 & -05:09:00.32 & 23.07 $\pm$ 0.02 
& 3862.83 & 2.1767 & b \\
NB387-C-07673 & 02:18:56.532 & -05:05:48.41 & 24.98 $\pm$ 0.07 
& 3853.55 & 2.1690 & c \\
NB387-C-08099 & 02:19:05.729 & -05:05:23.86 & 24.79 $\pm$ 0.05 
& 3889.02 & 2.1982 & c \\
NB387-C-08204 & 02:18:57.385 & -05:05:18.82 & 24.46 $\pm$ 0.05 
& 3895.63 & 2.2036 & b \\
NB387-C-08321 & 02:19:05.279 & -05:05:11.22 & 24.89 $\pm$ 0.06 
& 3891.62 & 2.2003 & b \\
NB387-C-09219 & 02:19:02.396 & -05:04:19.27 & 24.37 $\pm$ 0.04 
& 3890.03 & 2.1990 & c \\
NB387-C-09951 & 02:18:50.038 & -05:03:34.09 & 24.38 $\pm$ 0.04 
& 3901.78 & 2.2087 & b \\
NB387-C-11135 & 02:18:37.381 & -05:02:24.61 & 23.86 $\pm$ 0.03 
& 3882.13 & 2.1925 & b \\ 
NB387-C-12596 & 02:18:55.071 & -05:00:58.82 & 24.83 $\pm$ 0.06 
& 3886.29 & 2.1960 & c \\
NB387-C-16564 & 02:19:09.542 & -04:57:13.32 & 22.88 $\pm$ 0.03 
& 3861.44 & 2.1755 & a 
\enddata
\tablenotetext{(1)}{%
Coordinates are in J2000.
}
\tablenotetext{(2)}{%
NB387 magnitudes with $2\arcsec$ diameter apertures.
}
\tablenotetext{(3)}{%
Central wavelengths of observed Ly$\alpha$ lines (\AA). 
We perform a Gaussian fitting to each detected line 
using MPFIT, and derive the central wavelengths.
}
\tablenotetext{(4)}{%
Redshifts of Ly$\alpha$ lines estimated from the central 
wavelengths.
}
\tablenotetext{(5)}{%
Flags of reliability of the lines by inspecting the 2D spectra by eye: 
``a'' secure, ``b'' likely, ``c'' tentative.
}
\end{deluxetable*}

\begin{figure*}
\epsscale{0.74}
\plotone{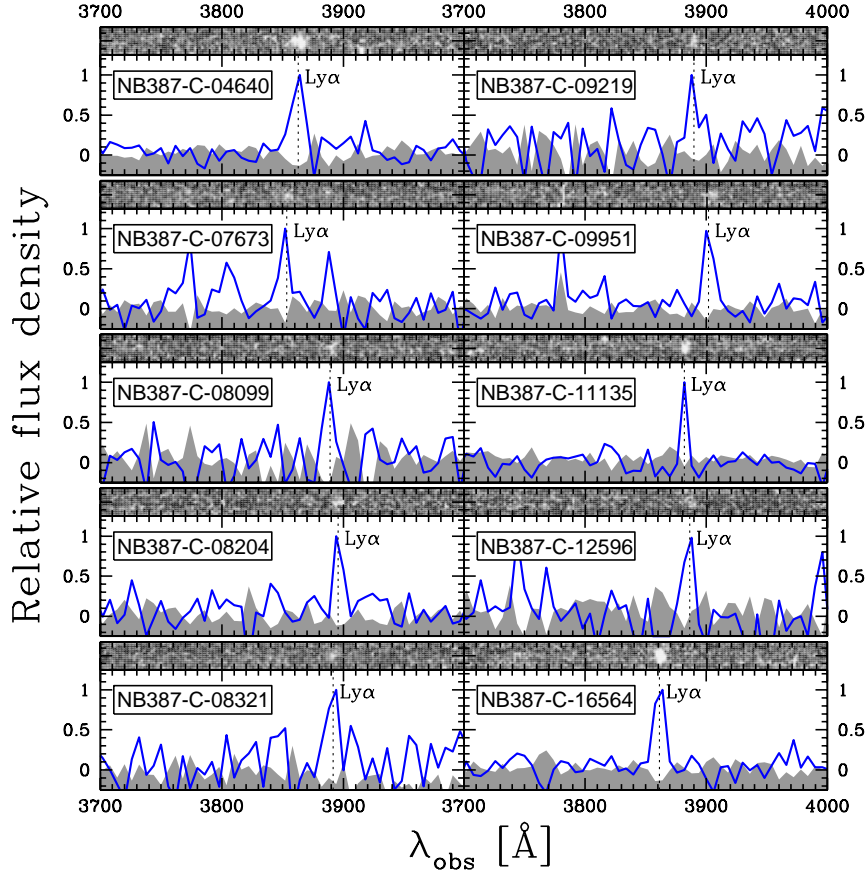}
\caption{
IMACS spectra of ten confirmed LAEs.
For each object, the top panel shows the $2D$ spectrum, while 
the bottom panel showing the $1$D spectrum (blue line)
and the sky background (gray shaded area),
both of which have been smoothed with a $3$ pixel boxcar filter
and arbitrarily normalized.
\label{fig:Lya_spectra}}
\end{figure*}

For Mg\,{\sc ii} and \CIV\ emitters, we use X-ray and radio imaging data,
since both emitters selected by Equation (\ref{eq:LAE_thresh})
should have large EWs (EW$_{\rm rest} \gtrsim30$\,\AA)
and thus they are likely to be AGNs.
For X-ray data, we use the \xmm\ $0.2 - 10.0$ keV band
source catalog by \citet{ueda2008}.
For radio data, we use the Very Large Array (VLA) $1.4$ GHz
source catalog by \citet{simpson2006}.
After removing some confused objects
by visual inspection, we find $11$ $(1)$ LAE candidates with
an X-ray (radio) counterpart.
The number of likely Mg\,{\sc ii} or \CIV\ emitters is thus $12$.
Note that $10$ out of the $12$ are also detected in the GALEX data.
\\

After removing spurious objects and obvious \OII\ and \CIV\ emitters,
we have $919$ $(=1,044-(21+90)-(12+12-10))$ LAE candidates.
Among them, $561$ are in the NB118 sub-region
and $105$ in the NB209 sub-region.
We plot all the candidates on the  $u^*-$NB387 vs. $B-$NB387 two color plane
in Figure \ref{fig:cc_diagram_data} and
on the $u^*-$NB387 vs. NB387 color-magnitude plane
in Figure \ref{fig:cm_diagram_data}.
Figure \ref{fig:cc_diagram_model} and Figure \ref{fig:cc_diagram_data}
show that the candidates (red circles) are placed in the isolated region
away from the locus of other galaxies and Galactic stars.

\subsection{Follow-up Spectroscopy Data} \label{ssec:specdata}

We carried out follow-up spectroscopy of $30$ objects
selected from the whole sample ($N=919$) so that
they are distributed in wide ranges of NB387 magnitudes 
and $u^*-$NB$387$ colors. 
The observations were made with the Inamori Magellan Areal
Camera and Spectrograph \citep[IMACS;][]{dressler2006}
on the Magellan telescope
using the $300$ lines mm$^{-1}$ grism and the WB$3600-5700$
filter on 2010 July 9-10 under photometric conditions.
We used the $f/4$ camera, which has a better sensitivity
than the $f/2$ camera at short wavelengths.
The total throughput of the $f/4$ camera with the $300$ lines mm$^{-1}$ 
grism at $3870$\AA\ is $6.2$\,\%, while that of the $f/2$ camera 
is $2.6$\,\%.
The on-source exposure time was $13,700$ seconds,
with a seeing size of $0\farcs 48 - 0\farcs 63$.
We chose a $0\farcs 8$ slit width, which
gives a resolving power of $R\sim 700$ around $4000$\,\AA.
The COSMOS pipeline was used for data reduction.

By inspecting the reduced spectra by eye, we detected
an emission line around $3870$\,\AA\ for ten objects,
while the remaining $20$ had no visible emission line.
The main reason for this low detection rate is
the bright limiting flux of our observation
due to the low sensitivity below $4000$\,\AA\
and the relatively short exposure time.
Indeed, all ten objects with line emission are brighter
than NB$387=25.0$, and the success rate limited to
NB$387<25.0$ ($13$ objects in total)
is found to be $10/13 = 77$\,\%.
Among the remaining three bright objects, two have 
NB387$\simeq 25.0$ and may be marginally undetected. 
The other one object has NB387 $= 24.4$, but smaller 
Ly$\alpha$ flux expected from its $u^*-$NB387 compared 
with the confirmed candidates.
The spectra of the ten confirmed LAEs are shown in 
Figure \ref{fig:Lya_spectra}, and the NB387 magnitudes and 
Ly$\alpha$-based redshifts are given in 
Table \ref{tbl:Lya_spectra}.

The ten objects are not \OII\ emitters at $z\simeq 0.04$
because of the lack of \OIII$\lambda 5007$ line at
the corresponding wavelength, $\simeq 5200$\,\AA.
They are not AGNs either, with Mg\,{\sc ii} emission at 
$z\simeq 0.4$ or \CIV\ emission at $z\simeq 1.5$, 
because of the absence of emission lines at $3600-5700$\,\AA\ 
(e.g., Mg\,{\sc ii}: \OII\ line at $\simeq 5150$\,\AA, 
\CIV: He\,{\sc ii}$\lambda1640$ line at $\simeq 4100$\,\AA\ and 
C\,{\sc iii}]$\lambda1909$ line at $\simeq 4770$\,\AA) 
and because we have removed AGN candidates from the sample in advance
(see \S \ref{ssec:interloper}).
Therefore, we conclude that all ten objects are LAEs
at $z\simeq 2.2$.
This demonstrates that the contamination in our LAE sample
is very low, at least for bright objects.

\section{\OII\ and \Ha\ Emission Lines} \label{sec:detections}

Our NIR images are not deep enough to study \OII\ and \Ha\
properties of LAEs based on individual detections.
Indeed, only $10$ objects have detection of \OII\ emission and
only $3$ have detection of \HaNII\ emission, as described in
\S \ref{ssec:detect_individual}.
We therefore carry out a stacking analysis of the whole LAE sample
in the NB118 sub-region,
and discuss average emission-line properties of $z\sim 2.2$ LAEs.
The individually detected objects are discussed below and 
compared with the stacked objects.

\subsection{Stacking Analysis} \label{ssec:detect_stacking}

Stacking is done separately for the NB118 sub-region and
the NB209 sub-region.
For the NB118 sub-region, we stack the NB118 and $J$ images
at the positions of $561$ LAE candidates.
Similarly, for the NB209 sub-region, the NB118, $J$,
NB209, and $K$ images are stacked at the positions of
the $105$ LAE candidates. 
Before stacking, 
we mask regions affected by a ghost, a stellar halo, and bad pixels, 
as well as regions with relatively large noise.
The stacked image in the NB118 sub-region are used to detect
the \OII\ flux at the highest $S/N$ ratio,
while the stacked image in the NB209 sub-region are used
to compare \OII\ and \Ha\ fluxes in a common sample.
To derive the average Ly$\alpha$ flux, we also stack NB387 and
$u^*$ images for each sub-region.
All stackings are done by median-stacking.
Before the stacking, we smooth the images with gaussian filters
so that both narrowband and broadband images have the same PSF sizes.
This enables us to measure the colors of the stacked objects 
by aperture photometry 
(\S \ref{sssec:photo_stacking}).
The PSF size before smoothing is $1\farcs 6$ for NB118 and $J$,
$1\farcs 2$ for NB209 and $K$, and $1\farcs 2$ for NB387 and $u^*$.
We do not remove the individually detected objects from 
the stacking analysis in order to increase the number of 
candidates for stackings.
In fact, results after removing the individually detected objects 
are consistent with the results in Equation (\ref{eq:color_stack}) 
owing to the median-stacking.

\begin{figure}
\epsscale{1.15}
\plotone{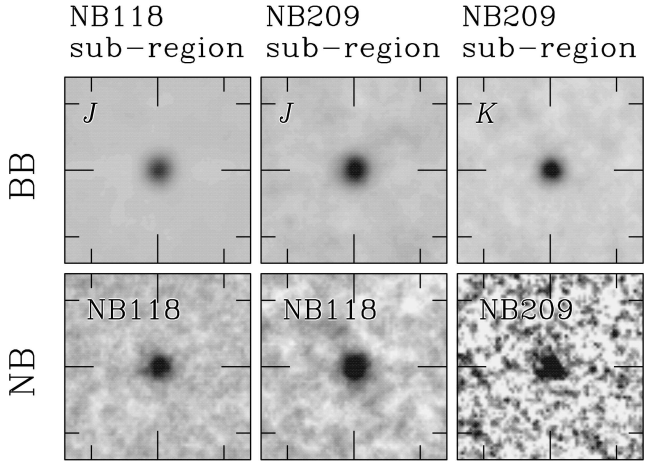}
\caption{
Snapshots of the stacked LAE in $J$ (top left) and NB$118$ (bottom left)
in the NB118 sub-region,
and $J$ (top center), NB$118$ (bottom center),
$K$ (top right), and NB$209$ (bottom right) in the NB209 sub-region.
Each image is $15\arcsec \times 15\arcsec$ in size.
\label{fig:stacked_image}}
\end{figure}

The stacked NIR images are shown in Figure \ref{fig:stacked_image}.
A signal is clearly visible in all the passbands
including NB118 and NB209.
We measure the magnitudes and colors of the stacked images
in \S \ref{sssec:photo_stacking}, and convert them into line fluxes
using Monte Carlo simulations in \S \ref{sssec:MCS}.

\subsubsection{Photometry} \label{sssec:photo_stacking}

\begin{deluxetable*}{llcccccc}
\tablecolumns{8}
\tabletypesize{\scriptsize}
\tablecaption{Properties of the stacked LAEs in the two sub samples
\label{tbl:data_stacked}}
\tablewidth{480pt}
\setlength{\tabcolsep}{0.11in}
\tablehead{
\colhead{sample } &
\colhead{Line} &
\colhead{$m^{B}_{\rm aper}$} &
\colhead{$m^{N}_{\rm aper}$} &
\colhead{$m^{N}_{\rm total}$} &
\colhead{EW$_{\rm rest}$} &
\colhead{Flux} &
\colhead{Luminosity} \\
\colhead{$\cdots$[$N$]\tablenotemark{(1)}} &
\colhead{} &
\colhead{(2)} &
\colhead{(3)} &
\colhead{(4)} &
\colhead{(5)} &
\colhead{(6)} &
\colhead{(7)} 
}
\startdata
NB118 sub  & \OII\ & $25.18\pm 0.02$ & $24.74 \pm 0.05$ & $24.57$ &
$106^{+14}_{-12}$ & $9.65^{+0.39}_{-0.39}\times 10^{-18}$ &
$3.54^{+0.14}_{-0.14}\times 10^{41}$ \\
$\cdots [561]$ & Ly$\alpha$ & $25.83\pm 0.01$ & $25.05 \pm 0.01$ & $24.87$ &
$86^{+3}_{-2}$ & $5.76^{+0.11}_{-0.09}\times 10^{-17}$ &
$2.11^{+0.04}_{-0.03}\times 10^{42}$ \\
\\
NB209 sub & \OII\ & $24.72\pm 0.03$ & $24.33 \pm 0.07$ & $24.11$ &
$96^{+23}_{-19}$ & $1.44^{+0.09}_{-0.10}\times 10^{-17}$ &
$5.26^{+0.35}_{-0.38}\times 10^{41}$ \\
$\cdots [105]$ & \HaNII\ & $24.72\pm 0.03$ & $24.07 \pm 0.10$ & $23.62$ &
$271^{+242}_{-104}$ & $2.18^{+0.34}_{-0.33}\times 10^{-17}$ &
$7.98^{+1.25}_{-1.21}\times 10^{41}$ \\
 & \Ha\tablenotemark{(8)} & \nodata & \nodata & \nodata & 
$256^{+229}_{-98}$ & $2.06^{+0.32}_{-0.31}\times 10^{-17}$ & 
$7.55^{+1.19}_{-1.15}\times 10^{41}$ \\
 & Ly$\alpha$ & $25.75\pm 0.02$ & $25.09 \pm 0.02$ & $24.94$ &
$63^{+3}_{-5}$ & $4.90^{+0.12}_{-0.19}\times 10^{-17}$ &
$1.80^{+0.05}_{-0.07}\times 10^{42}$
\enddata
\tablenotetext{(1)}{%
Numbers of stacked objects in square brackets.
}
\tablenotetext{(2)}{%
Broadband aperture magnitudes and their $1\sigma$ errors.
Broadbands are $J$ for \OII, $K$ for \HaNII,
and $u^*$ for Ly$\alpha$.
The diameter of the aperture is $2\farcs 5 - 3\farcs 2$.
Aperture sizes are chosen to have twice the PSF sizes.
}
\tablenotetext{(3)}{%
Narrowband aperture magnitudes and their $1\sigma$ errors.
Narrowbands are NB118 for \OII, NB209 for \HaNII,
and NB387 for Ly$\alpha$.
The diameter of the aperture is $2\farcs 5 - 3\farcs 2$.
Aperture sizes are chosen to have twice the PSF sizes.
}
\tablenotetext{(4)}{%
Narrowband total magnitudes.
The diameters of the aperture are $4\farcs 8$ and $6\farcs 1$ 
for \OII\ and Ly$\alpha$ line in the NB118 sub-region, respectively, 
and $5\farcs 7$, $5\farcs 7$, and $6\farcs 1$ 
for \OII, \HaNII, and Ly$\alpha$ line in the NB209 sub-region, 
respectively.
Aperture sizes are chosen to include close to $100$\,\% 
of the flux.
}
\tablenotetext{(5)}{%
Rest-frame equivalent width of the lines (\AA).
For \OII\ and \HaNII, EWs are estimated by Monte Carlo simulations
(see \S \ref{sssec:MCS} in details),
and for Ly$\alpha$, the colors of $u^*-$NB387 are used for
the estimates of EWs.
}
\tablenotetext{(6)}{%
\uppercase{}Line fluxes in unit of erg\,s$^{-1}$\,cm$^{-2}$.
}
\tablenotetext{(7)}{%
\uppercase{}Line luminosities in unit of erg\,s$^{-1}$.
}
\tablenotetext{(8)}{%
The contribution from \NII$\lambda\lambda6584,6548$ lines is 
subtracted from the observed \HaNII\ luminosity using the metallicity 
estimated in \S \ref{sssec:determin_Z} (see also \S \ref{sssec:SFR_derive}).
}
\end{deluxetable*}

We measure aperture magnitudes of the stacked objects
using the IRAF task \verb+phot+.
The aperture diameter is set to be $2\farcs 5 - 3\farcs 2$ 
(about twice the PSF size) to calculate colors,
while larger apertures of $4\farcs 8 - 6\farcs 1$ (depending on the passband)
are adopted to obtain total magnitudes.
Errors in the magnitudes are estimated
in the same manner as in \citet{ono2010b};
we create $1,000$ median-stacked sky noise images,
each of which is made of $561$ ($105$) randomly-selected
sky noise images in the NB118 (NB209) sub-regions.
We then fit the negative part of histograms with a Gaussian,
whose FWHMs are used to estimate the limiting magnitudes.
The magnitudes and errors obtained are summarized
in Table \ref{tbl:data_stacked}.
The uncertainties of zero points are not included in the errors.

\subsubsection{Equivalent Widths and Fluxes of the Lines} \label{sssec:MCS}

The $J-$NB118 and $K-$NB209 colors
for the stacked objects are calculated to be:
\begin{eqnarray}
&& J-{\rm NB118} ({\rm NB118 sub}) = 0.43 \pm 0.05 \nonumber \\
&& J-{\rm NB118} ({\rm NB209 sub}) = 0.39 \pm 0.08 ,  \label{eq:color_stack} \\
&& K-{\rm NB209} ({\rm NB209 sub}) = 0.65 \pm 0.11 \nonumber
\end{eqnarray}
where the color with 'NB118sub' and 'NB209sub' in round brackets
is derived from the stacked object in the NB118 sub-region
and that in the NB209 sub-region, respectively. 
These large, positive values with the small errors indicate
significant detection of the lines.
In order to estimate the \OII\ and \HaNII\ equivalent widths and fluxes 
with the best accuracy, we used Monte Carlo simulations. 
If both a narrowband and a broadband (which brackets the narrowband)
filters have ideal top-hat response functions,
an EW of an emission line falling in the narrowband can be
calculated as:
\begin{eqnarray}
{\rm EW}_{\rm rest} = \frac{\left( f^{N}_{\lambda}-f^{B}_{\lambda}\right)
      \Delta \lambda^N \Delta \lambda^B}{\Delta \lambda^B f^{B}_{\lambda}
      - \Delta \lambda^N f^{N}_{\lambda}}\frac{1}{1+z} , 
      \label{eq:EW_0th}
\end{eqnarray}
where $f_{\lambda}$ is the flux density per unit wavelength,
$\Delta \lambda$ is the width of a given filter, $z$ is redshift, and
superscripts $N$ and $B$ indicate narrowband and broadband, respectively.
To derive this formula, we have assumed that the flux density
of the continuum emission is constant over the whole wavelength
range.
If this formula is used, the colors obtained above
are converted into
EW$_{\rm rest, NB118 sub}$(\OII)$= 25$\,\AA ,
EW$_{\rm rest, NB209 sub}$(\OII)$= 23$\,\AA , and
EW$_{\rm rest, NB209 sub}$(\HaNII)$= 58$\,\AA, respectively.

\begin{figure*}
\epsscale{1.15}
\plotone{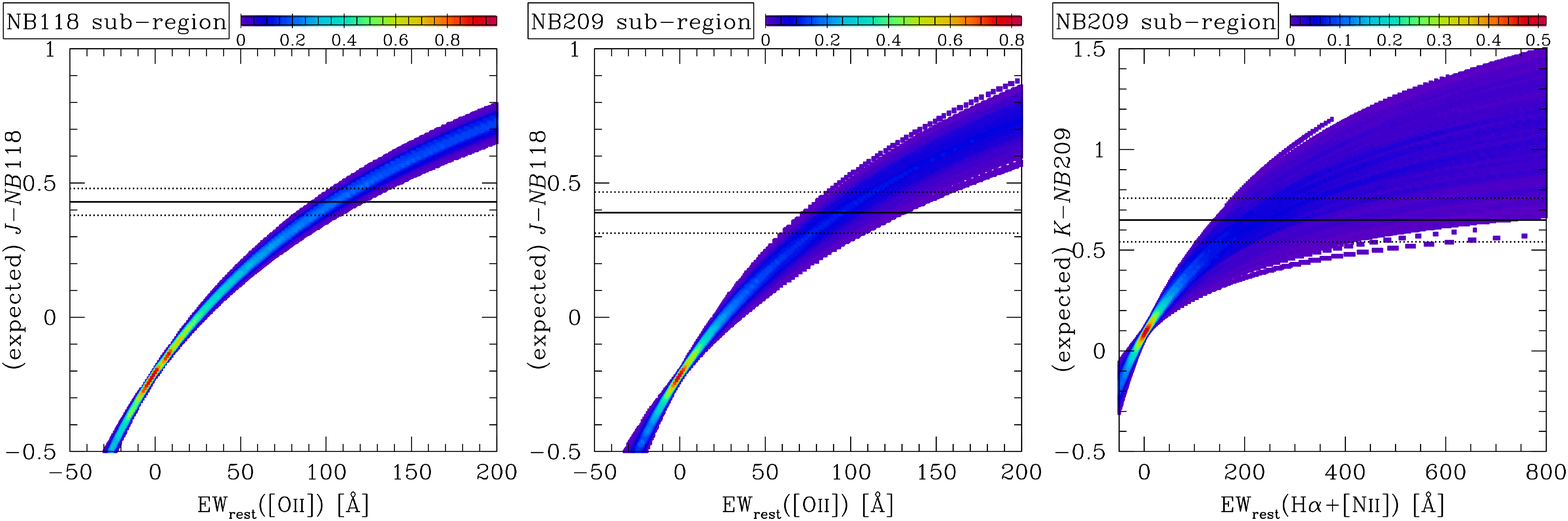}
\caption{
Relation between EW$_{\rm rest}$ and expected $J-$NB118
and $K-$NB209 colors from the Monte Carlo simulations:
(left) EW$_{\rm rest}$(\OII) vs. $J-$NB118 in the NB118 sub-region;
(center) EW$_{\rm rest}$(\OII) vs. $J-$NB118 in the NB209 sub-region;
(right) EW$_{\rm rest}$(\HaNII) vs. $K-$NB209 in the NB209 sub-region.
The color contours indicate the probability of
$J-$NB118 or $K-$NB209 at given EW$_{\rm rest}$;
redder colors mean higher possibilities as shown in the color bars in the top.
The solid black lines correspond to
the measured $J-$NB118 or $K-$NB209 colors and
the dotted black lines are their $1\sigma$ errors. 
\label{fig:MCS_J_K}}
\end{figure*}

Although these values are useful as zero-order estimates,
their accuracy is not sufficient for our purpose. 
In fact the assumption on the shape of the passband used to derive 
Equation (\ref{eq:EW_0th}) is over simplified in two aspects: 
the actual NB118 and NB209 passbands are not top-hat
but rather close to a triangle shape
and the central wavelengths of these passbands vary over
the FoV (Equation (\ref{eq:NIR_NB_shift}); see \S \ref{ssec:NIR_NB}).
These issues must be considered to obtain correct EW values,
since the objects used for stacking have different redshifts
(corresponding to the different locations within the band 
width of NB387) and they are distributed across the FoV.
In general, Equation (\ref{eq:EW_0th}) is correct only when
a line is located where the response function of
the narrowband peaks.
In reality, however, a large fraction of the LAEs are expected to have 
\OII\ and \HaNII\ lines off the peak responses of NB118 and NB209 
based on the re-scaled shapes of NB387 (see Figure \ref{fig:filters_all}).
This means that the EWs of stacked objects calculated
by Equation (\ref{eq:EW_0th}) are always underestimated.

We carry out Monte Carlo simulations taking into account of
the above two points more accurately to estimate the EWs from
the observed colors of the stacked objects.
As an example, we describe below the simulations for the \OII\ 
line of the objects in the NB118 sub-region.
The simulations for \OII\ and \HaNII\ lines in the NB209 sub-region
are essentially the same.

The simulations are carried out under the assumption that
all $561$ objects have identical spectra, i.e., the same EW
and the same underlying continuum spectra.
For the continuum spectra, we use the best-fit spectrum
from the SED fitting to the stacked LAE (see \S\ref{sec:SED_fitting})%
\footnote{
We also try two independent SEDs for the simulation; the best-fit SED 
of spectroscopically confirmed BX galaxies at $2.0<z<2.2$ with 
constant star formation history \citep{guaita2011} and 
the best-fit spectrum from the SED fitting to the $z=3.1$ stacked LAE
\citep{ono2010a}, and find that the results are consistent with each other 
well within their $1\sigma$ errors.
}.
We then vary EW$_{\rm rest}$ over $-100$\,\AA\ and $300$\,\AA\ with
1\,\AA\ interval,
and for each value we carry out a Monte Carlo simulation
described by steps 1 -- 4 below to derive the probability 
distribution of $J-$NB118 color for that EW.
We thus simulate the relation between EW and $J-$NB118.
The EW of the real, stacked object is calculated
by substituting the observed $J-$NB118 color into the relation.

\begin{enumerate}
\item
We first generate $561$ LAEs with a given EW value,
and assign to each one of the $561$ positions on the FoV
of the real objects without duplication.
Then, assuming that the (Ly$\alpha$) redshift distribution of our LAEs
is same as the shape of the NB387 response function,
with a peak at $z=2.18$ and an FWHM of $\Delta z = 0.075$,
we randomly select a redshift for each of the $561$ objects
from this distribution.

\item
Spectroscopic observations of high-redshift star-forming
galaxies have found that the redshift measured from Ly$\alpha$ line
is offset from the nebular-line redshift
by a few hundred km\,s$^{-1}$, due to radial acceleration of 
the circumgalactic gas emitting the line 
(e.g., \citealt{pettini2001,steidel2010,mclinden2011,finkelstein2011}).
We randomly assign a redshift offset to each of the $561$ objects
simulated in step 1,
assuming that the offsets obey the distribution function obtained
by \citet{steidel2010} for $z\sim 2$ galaxies
\footnote{
Although LAEs are so far found to have systematically 
smaller velocity offsets than LBGs, 
$\sim 150$\,km\,s$^{-1}$ for LAEs \citep{mclinden2011,finkelstein2011}  
while $\sim 400$\,km\,s$^{-1}$ for LBGs \citep{steidel2010}, 
we use the data of \citet{steidel2010} 
since it is based on much a larger number of measurements 
($>40$ LBGs while $4$ LAEs) and thus statistically more reliable.
In any case, the wavelength shift caused by velocity offset is 
much smaller than that due to the variation of the response curve 
over the FoV  
($\sim 15$\AA\ by a velocity offset of $\sim 400$\,km\,s$^{-1}$ 
while up to $\sim 150$\AA\ by the positional variation for NB118). 
}.
Each object is thus given an \OII\ redshift.

\item
For each object we calculate $f_\lambda^J$ and $f_\lambda^{\rm NB118}$
using the EW, continuum spectrum, \OII\ redshift,
and the NB118 response function at the position of the object
on the FoV.
We then sum the $561$ objects' $f_\lambda^J$ and $f_\lambda^{\rm NB118}$,
and divide the former by the latter to obtain the $J-$NB118 color of
the stacked object.

\item
We repeat steps 1 -- 3 $500$ times to obtain the probability
distribution of $J-$NB118 color for the given EW.

\end{enumerate}

\begin{deluxetable*}{lcccccccc}
\tablecolumns{9}
\tabletypesize{\scriptsize}
\tablecaption{Properties of individual objects %
\label{tbl:data_OII_Ha_Lya}}
\tablewidth{0pt}
\setlength{\tabcolsep}{0.05in}
\tablehead{
\colhead{ID} &
\colhead{R.A.\tablenotemark{(1)}} &
\colhead{Dec.\tablenotemark{(1)}} &
\colhead{EW$_{{\rm Ly}\alpha}$} &
\colhead{$L_{{\rm Ly}\alpha}$} &
\colhead{EW$_{{\rm [OII]}}$} &
\colhead{$L_{{\rm [OII]}}$}  &
\colhead{EW$_{{\rm H}\alpha+{\rm [NII]}}$} &
\colhead{$L_{{\rm H}\alpha+{\rm [NII]}}$} \\
\colhead{} &
\colhead{} &
\colhead{} &
\colhead{(2)} &
\colhead{(3)} &
\colhead{(4)} &
\colhead{(5)}  &
\colhead{(6)} &
\colhead{(7)}
}
\startdata
NB387-C-06908 & 02:17:57.113 & -05:06:37.06 & $45^{+9}_{-7}$ & $4.58^{+0.71}_{-0.65}$ & $277^{+154}_{-93}$ & $3.90^{+0.17}_{-0.22}$ & \nodata & \nodata \\
NB387-C-22326 & 02:17:11.390 & -04:51:14.10 & $54^{+23}_{-17}$ & $1.56^{+0.49}_{-0.42}$ & \nodata & \nodata & $198^{+67}_{-55}$ & $2.93^{+0.28}_{-0.34}$ \\
NB387-C-26967 & 02:17:19.875 & -04:46:54.76 & $31^{+13}_{-10}$ & $0.96^{+0.33}_{-0.28}$ & $22^{+10}_{-8}$ & $2.27^{+0.56}_{-0.65}$ & \nodata & \nodata \\
NB387-N-01923 & 02:17:55.840 & -04:47:40.11 & $121^{+32}_{-24}$ & $4.87^{+0.75}_{-0.68}$ & $127^{+32}_{-26}$ & $3.81^{+0.18}_{-0.21}$ & \nodata & \nodata \\
NB387-S-25047 & 02:17:32.417 & -05:12:51.09 & $31^{+3}_{-2}$ & $19.43^{+1.39}_{-1.35}$ & $18^{+2}_{-2}$ & $6.90^{+0.48}_{-0.50}$ & \nodata & \nodata \\
NB387-W-00165 & 02:17:07.486 & -04:53:53.75 & $59^{+16}_{-13}$ & $3.45^{+0.72}_{-0.64}$ & $35^{+19}_{-15}$ & $1.40^{+0.32}_{-0.39}$ & \nodata & \nodata \\
NB387-W-00372 & 02:17:04.957 & -04:45:35.62 & $32^{+7}_{-6}$ & $4.94^{+0.95}_{-0.87}$ & $20^{+9}_{-7}$ & $1.63^{+0.40}_{-0.45}$ & \nodata & \nodata \\
NB387-W-02225 & 02:16:46.049 & -04:59:05.44 & $39^{+12}_{-10}$ & $1.82^{+0.44}_{-0.40}$ & $10^{+8}_{-7}$ & $0.86^{+0.45}_{-0.53}$ & \nodata & \nodata \\
NB387-W-04041 & 02:16:28.164 & -04:45:17.72 & $63^{+18}_{-14}$ & $2.99^{+0.62}_{-0.55}$ & $75^{+19}_{-16}$ & $3.87^{+0.27}_{-0.31}$ & $406^{+66}_{-58}$ & $6.80^{+0.20}_{-0.22}$ \\
NB387-W-04492 & 02:16:23.751 & -04:57:57.92 & $43^{+13}_{-11}$ & $2.31^{+0.55}_{-0.49}$ & $92^{+37}_{-28}$ & $2.98^{+0.26}_{-0.32}$ & $68^{+27}_{-24}$ & $2.08^{+0.43}_{-0.51}$ \\
NB387-W-06136 & 02:16:07.936 & -05:00:07.96 & $40^{+8}_{-7}$ & $5.50^{+0.94}_{-0.87}$ & $12^{+3}_{-3}$ & $2.44^{+0.47}_{-0.50}$ & \nodata & \nodata 
\enddata
\tablenotetext{(1)}{%
Coordinates are in J2000.
}
\tablenotetext{(2)}{%
Rest-frame equivalent width of Ly$\alpha$ in unit of \AA.
}
\tablenotetext{(3)}{%
\uppercase{}Ly$\alpha$ luminosity in unit of $10^{42}$\,erg\,s$^{-1}$.
}
\tablenotetext{(4)}{%
Rest-frame equivalent width of \OII\ in unit of \AA.
}
\tablenotetext{(5)}{%
\OII\ luminosity in unit of $10^{42}$\,erg\,s$^{-1}$.
}
\tablenotetext{(6)}{%
Rest-frame equivalent width of \HaNII\ in unit of \AA.
}
\tablenotetext{(7)}{%
\HaNII\ luminosity in unit of $10^{42}$\,erg\,s$^{-1}$.
}
\end{deluxetable*}

Figure \ref{fig:MCS_J_K} shows the results of
the simulations, i.e., the distribution of $J-$NB118 and $K-$NB209
color as a function of EW.
In each panel, the color contour indicates the probability of 
$J-$NB118 or $K-$NB209 at given EWs, and 
the solid line and two dotted lines indicate
the central value and the $1\sigma$ lower and upper limits,
respectively, of the observed color.
We translate the observed colors and their errors 
into EWs in the following manner. 
First, we assume that for each of $J-$NB118 and $K-$NB209, 
the probability distribution of the true value
is a Gaussian with its $\sigma$ equal to the observed $1\sigma$ 
error in that color.
Next, we randomly select a color following this Gaussian 
probability distribution, and assign to it an EW based on the EW 
probability distribution against color shown in Figure \ref{fig:MCS_J_K}.
We repeat this procedure $10,000$ times. 
Finally, we sort the $10,000$ EWs in ascending order, 
and regard the $50$\,\%-tile, $16\%$-tile, and $84$\,\%-tile as 
the central value, $1\sigma$ lower limit, and $1\sigma$ upper limit, 
respectively.
From the translation procedures, we obtain EW$_{\rm rest}$ for 
\OII\ and \HaNII: 
\begin{eqnarray}
&& {\rm EW}_{\rm rest, NB118 sub}({\rm [OII]}) = 106^{+14}_{-12}
\,{\rm \AA} \nonumber \\
&& {\rm EW}_{\rm rest, NB209 sub}({\rm [OII]}) = 96^{+23}_{-19}
\,{\rm \AA}. \label{eq:EW_stack} \\
&& {\rm EW}_{\rm rest, NB209 sub}({\rm H}\alpha+{\rm [NII]}) = 271^{+242}_{-104}
\,{\rm \AA} \nonumber
\end{eqnarray}
The errors in EW correspond to $1\sigma$.
The large scatter in EW$_{\rm rest, NB209 sub}$(\HaNII)
is due to the large photometric error of $K-$NB209
(see Equation (\ref{eq:color_stack})) and
the large variance of the NB209 passband over the FoV.

We calculate the \OII\ and \Ha\ fluxes from
the EWs obtained here combined with the total magnitudes derived in
\S \ref{sssec:photo_stacking}, and convert them into luminosities.
These values are summarized in Table \ref{tbl:data_stacked}.

We estimate the Ly$\alpha$ EW in a similar manner to what 
we explain above; 
we run the same Monte Carlo simulation skipping step $2$, 
and then translate the $u^*-$NB387 color into the Ly$\alpha$ EW.
The Ly$\alpha$ EW$_{\rm rest}$ derived for the NB118 (NB209) sub-region is 
$86^{+3}_{-2}$ ($63^{+3}_{-5}$) \AA, from which we obtain 
the observed Ly$\alpha$ luminosity to be $2.11^{+0.04}_{-0.03}\times 10^{42}$ 
($1.80^{+0.05}_{-0.07}\times 10^{42}$) erg\,s$^{-1}$ 
(Table \ref{tbl:data_stacked}).

From previous narrow-band surveys, 
LAEs with $L$(Ly$\alpha)= 10^{42} - 10^{44}$ erg\,s$^{-1}$ 
are referred to as typical LAEs (e.g., \citealt{gronwall2007,ouchi2008})
Therefore, the stacked LAEs we obtain are expected to have 
average properties of LAEs with $L$(Ly$\alpha)>10^{42}$.

\subsection{Individual Objects} \label{ssec:detect_individual}

Seventeen out of the $561$ LAE candidates in the NB118 sub-region
are detected in NB118 at $\ge 5 \sigma$ levels,
among which ten have $J-$NB$118 \ge 0.0$ and
${\rm NB118}<{\rm NB118}\,(5\sigma)$%
\footnote{The $5\sigma$ limiting magnitude here is defined 
with aperture sizes of $3\farcs 2$ diameter for NB118 and 
$2\farcs 5$ diameter for NB209, 
and estimated to be $\simeq 22.9$ and $\simeq 22.4$, respectively.
}, 
indicative of the presence of \OII\ emission. 
A detailed calculation using the continuum spectrum from the SED fitting
shows that these detection criteria correspond to
EW$_{\rm rest}($\OII$) \gtrsim 7$\,\AA\ and
$f($\OII$) \gtrsim 1.2\times 10^{-17}$ erg\,s$^{-1}$\,cm$^{-2}$.
Similarly, seven out of the $105$ objects in the NB209 sub-region are
detected in NB209 at $\ge 5 \sigma$ levels, among which three has
$K-$NB$209>0.3$ and ${\rm NB209}<{\rm NB209}\, (5\sigma)$, 
indicative of the presence of \HaNII\ emission 
at more than the $5\sigma$ level. 
These detection criteria correspond to
EW$_{\rm rest}($\HaNII$) \gtrsim 22$\,\AA\ and
$f$(\HaNII$) \gtrsim 2.1\times 10^{-17}$ erg\,s$^{-1}$\,cm$^{-2}$.
Two of the \HaNII\ detected objects (NB387-W-04041 and NB387-W-04492) 
are also detected in \OII, 
while one (NB387-C-22326) is not detected in \OII, 
probably due to a variation of dust extinction and 
the complicated origin of the \OII\ luminosity 
(see \S \ref{sssec:SFR_derive_OII}).
The EWs and line luminosities of these objects are summarized in
Table \ref{tbl:data_OII_Ha_Lya}.

\subsection{Average EWs and Luminosities of \OII\ and \Ha} 
\label{ssec:properties_oii_ha}

Before presenting the physical properties derived from \OII\ 
and \Ha\ emission lines, we summarize the strengths 
of these lines of the stacked LAEs.

The stacked object from the NB118 sub-region has 
EW(\OII) $= 106^{+14}_{-12}$\,\AA\ 
and 
$L$(\OII)$ = 3.54^{+0.11}_{-0.11}\times 10^{41}$\,erg\,s$^{-1}$.
This EW(\OII) is much larger than those obtained for
typical high-$z$ galaxies 
(e.g., $z\sim 1$ galaxies from the DEEP2 survey; \citealt{cooper2006}).

The stacked object from the NB209 sub-region has 
EW(\HaNII) = $271^{+242}_{-104}$\,\AA\ 
and 
$L$(\HaNII) = $7.98^{+1.17}_{-1.15}\times 10^{41}$\,erg\,s$^{-1}$, 
which correspond to 
EW(\Ha) = $256^{+229}_{-98}$\,\AA\ and 
$L$(\Ha) = $7.55^{+1.11}_{-1.10}$\,erg\,s$^{-1}$ 
after subtraction of \NII\ emission (see \S \ref{sssec:SFR_derive}).
This EW(\Ha) is larger than those 
obtained for other high-$z$ galaxies; 
e.g., \citet{erb2006b} performed \Ha\ spectroscopy of
$z\sim 2$ UV-selected galaxies to find the median EW(\Ha) $\sim 170$\AA\
(Note that EW(\Ha) of UV-selected galaxies from \Ha\ spectroscopy 
may be biased high, since objects with stronger \Ha\ emission 
can be observed more easily).
Recently, \citet{cowie2011} have found that the bulk ($\sim 75$\,\%) 
of the local LAEs have EW(\Ha) $>100$\,\AA. 
Thus, it seems that a large \Ha\ EW is a common character of LAEs 
irrespective of redshift.

\section{SED Fitting} \label{sec:SED_fitting}

\begin{figure*}
\epsscale{1.00}
\plotone{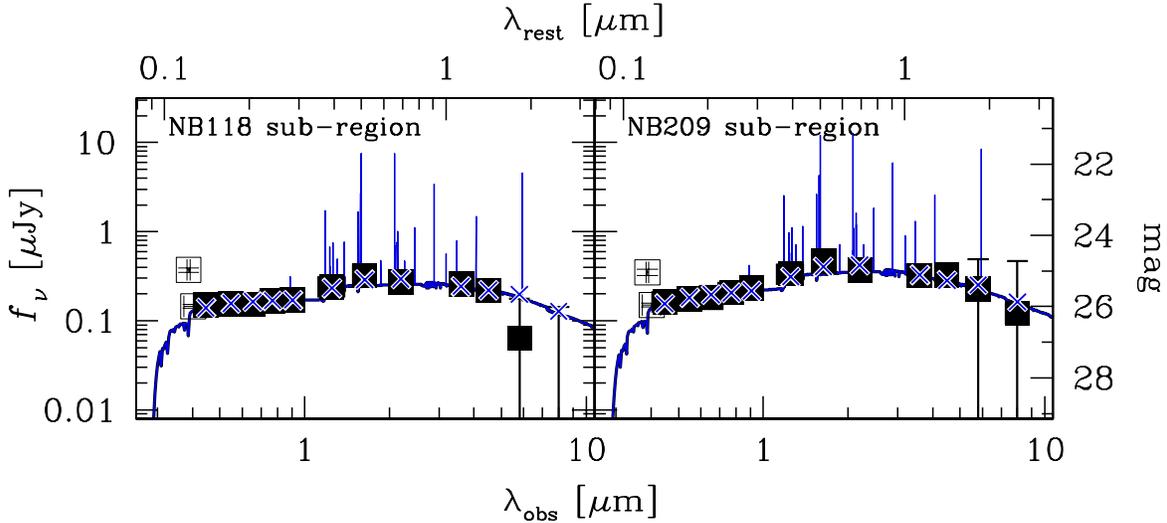}
\caption{
Results of SED fitting for the stacked objects in the NB118
sub-region (left) and the NB209 sub-region (right).
The filled squares show the observed flux densities used
for the fitting ($B,V,R,i',z',J,H,K,[3.6],[4.5],[5.8],$
and $[8.0]$), while the open squares indicate those
not used for the fitting ($u^*$ and NB387).
The blue lines show the best-fit model spectra,
and the blue crosses correspond to the best-fit flux densities.
\label{fig:SED_fitting}}
\end{figure*}

\begin{deluxetable*}{lcccccccccccc}
\tablecolumns{13}
\tabletypesize{\scriptsize}
\tablecaption{Broadband Photometry of the stacked LAEs for SED fitting
\label{tbl:photometry_SEDfit}}
\tablewidth{0pt}
\setlength{\tabcolsep}{0.05in}
\tablehead{
\colhead{sample} &
\colhead{$B$} &
\colhead{$V$} &
\colhead{$R$} &
\colhead{$i'$} &
\colhead{$z'$} &
\colhead{$J$} &
\colhead{$H$} &
\colhead{$K$} &
\colhead{$[3.6]$} &
\colhead{$[4.5]$} &
\colhead{$[5.8]$} &
\colhead{$[8.0]$} 
}
\startdata
NB118 sub-region 
& $25.96$
& $25.93$
& $25.95$
& $25.85$
& $25.82$
& $25.44$
& $25.12$
& $25.30$
& $25.38$
& $25.56$
& $26.90$
& $99.99$ \\
& $(29.31)$
& $(29.28)$
& $(29.29)$
& $(29.20)$
& $(29.13)$
& $(28.60)$
& $(28.06)$
& $(28.41)$
& $(28.56)$
& $(28.08)$
& $(26.26)$
& $(26.00)$ \\ 
\\
NB209 sub-region
& $25.87$
& $25.80$
& $25.75$
& $25.60$
& $25.48$
& $25.09$
& $24.72$
& $24.96$
& $25.13$
& $25.11$
& $25.51$
& $26.19$ \\
& $(29.20)$
& $(29.11)$
& $(29.07)$
& $(28.93)$
& $(28.73)$
& $(28.06)$
& $(27.38)$
& $(27.86)$
& $(27.62)$
& $(27.28)$
& $(25.35)$
& $(25.04)$
\enddata
\tablecomments{
Broadband photometry of the stacked objects 
whose stackings are performed for objects with 
IRAC coverage (see \S\ref{sec:SED_fitting}).
All magnitudes are total magnitudes.
99.99 mag means negative flux densities.
Magnitudes in parentheses are $1\sigma$ uncertainties 
adopted in SED fitting. 
}
\end{deluxetable*}

\begin{deluxetable*}{lcccccc}
\tablecolumns{7}
\tabletypesize{\scriptsize}
\tablecaption{Physical Properties of the stacked objects from SED fitting
\label{tbl:properties_SEDfit}}
\tablewidth{380pt}
\setlength{\tabcolsep}{0.13in}
\tablehead{
\colhead{sample} &
\colhead{$f_{\rm esc}^{\rm ion}$} &
\colhead{$Z$} &
\colhead{$M_{\star}$} &
\colhead{$E(B-V)_{\star}$} &
\colhead{Age} &
\colhead{$\chi^2_r$} \\
\colhead{} &
\colhead{} &
\colhead{[$Z_{\odot}$]} &
\colhead{[$10^8 M_{\odot}$]} &
\colhead{[mag]} &
\colhead{[Myr]} &
\colhead{} 
}
\startdata
NB118 sub-region 
& $0.7^{+0.1}_{-0.3}$
& $0.2$
& $2.88^{+0.43}_{-0.13}$
& $0.21^{+0.02}_{-0.04}$
& $12.6^{+17.6}_{-2.6}$
& $1.678$ \\
\\
NB209 sub-region
& $0.8^{+0.0}_{-0.1}$
& $0.2$
& $4.79^{+0.22}_{-0.81}$
& $0.27^{+0.01}_{-0.03}$
& $8.32^{+1.68}_{-1.40}$
& $1.661$
\enddata
\tablecomments{
Derived physical properties and their $1\sigma$ errors 
of the stacked objects from SED fitting.
$Z$ is fixed to $0.2Z_{\odot}$.
The degree of freedom is $8$.
The results may suffer from additional errors 
due to degeneracy with age or stellar metallicity, 
and due to possible systematic uncertainties.
}
\end{deluxetable*}

We perform SED fitting for the stacked objects 
in the NB118 sub-region and the NB209 sub-region to infer 
their stellar populations.
We note that the stacked objects used for the SED fitting are not 
exactly the same as those presented in \S \ref{ssec:detect_stacking}, 
but the stacking is performed only for objects with {\it Spitzer}/IRAC 
$3.6, 4.5, 5.8,$ and $8.0\mu$m photometry from the {\it Spitzer} 
legacy survey of the UDS field (SpUDS; {\it Spitzer} Proposal ID 40021; 
PI: J. Dunlop) so that stellar population parameters be well constrained%
\footnote{
We demand only IRAC coverage, not IRAC detection.
}. 
After removing objects with any confusion from neighboring objects 
in the IRAC images by eye, 
we stack $304$ and $55$ LAE candidates in the NB118 sub-region and 
the NB209 sub-region, respectively.

The procedure of the SED fitting is the same as 
that of \citet{ono2010b}, except for a fixed redshift ($z=2.18$).
We use the stellar population synthesis model GALAXEV \citep{BC2003} 
for stellar SEDs, and include nebular emission \citep{SdB2009}.
A Salpeter initial mass function (IMF; \citealt{salpeter1955}) is assumed. 
We choose constant star formation history 
and the stellar metallicity $Z=0.2\,Z_{\odot}$ since previous studies 
have shown that most LAEs are young, and that constant star formation 
history and subsolar stellar metallicities are reasonable assumptions.
Since we include nebular emission in the fitting, 
varying star formation history and stellar metallicity 
makes the fitting too complicated.
\citet{guaita2011} performed SED fitting to $z\sim 2.1$ stacked LAEs 
with three scenarios of star formation histories, 
exponentially increasing, decreasing, and constant star formation, 
to find equally good fits to the data.
They also note that among the free parameters 
SED fitting can relatively well constrain stellar mass and dust extinction, 
which are of particular interest in this paper.
We also note that the assumption of subsolar metallicity 
is consistent with the gas-phase metallicity derived for our stacked LAE 
(see \S \ref{ssec:dis_metallicity}).
For dust extinction, we use Calzetti's extinction law 
\citep{calzetti2000} on the assumption of  $E(B-V)_{\rm gas}=E(B-V)_{\star}$ 
as proposed by \citet{erb2006b}.
IGM attenuation is calculated using the prescription given by 
\citet{madau1995}. 
We do not use $u^*$ and NB387 data since they are significantly contaminated 
by strong Ly$\alpha$ emission.
Table \ref{tbl:photometry_SEDfit} summarizes the broadband photometry 
of the stacked objects that are used for SED fitting.
The uncertainties in optical and NIR band photometry listed in
Table \ref{tbl:photometry_SEDfit} contain photometric and 
two systematic errors. The photometric errors are estimated in the 
same manner as explained in \S \ref{sssec:photo_stacking},
and the two systematic errors, associated with aperture correction and
zero point, are given as follows; 
the errors in aperture correction are estimated to be $0.01-0.03$ mag 
using point sources, while the zero-point uncertainties are assumed 
to be $0.05$ mag for every optical and NIR band (inferred by 
\citet{furusawa2008} for optical bands). The photometric and 
two systematic errors are added in quadrature. 
For the IRAC channels, we include only photometric errors and do not 
include any systematic errors since they are unknown and 
the photometric errors are large enough to be dominant uncertainties. 
Inclusion of systematic errors to the IRAC channels will not change 
the results of the SED fitting significantly. 
More details of the SED fitting will be reported elsewhere 
(Y. Ono et al. in preparation).

We obtain 
$M_{\star}=2.88^{+0.43}_{-0.13}\times 10^{8}\,M_{\odot}$, 
$E(B-V)=0.21^{+0.02}_{-0.04}$,
and age $=1.26^{+1.76}_{-0.26}\times 10^{7}$\,yr for the stacked object 
in the NB118 sub-region, with a reduced chi-squares of 
$\chi^2_r=1.68$, and 
$M_{\star}=4.79^{+0.22}_{-0.81}\times 10^{8}\,M_{\odot}$, 
$E(B-V)=0.27^{+0.01}_{-0.03}$, 
and age $=8.32^{+1.68}_{-1.40}\times 10^{6}$\,yr for the stacked object 
in the NB209 sub-region, with $\chi^2_r=1.66$. 
Table \ref{tbl:properties_SEDfit} summarizes the physical properties 
derived from the SED fitting, and  
Figure \ref{fig:SED_fitting} shows the best-fit model spectra 
with the observed flux densities.
The errors in the best-fit parameters correspond to the $1\sigma$ 
confidence interval ($\Delta \chi^2<1$) for each parameters. 
The relatively small errors of the results are due to the small 
uncertainties adopted in the SED fitting. We note that the results
from the SED fitting may suffer from additional errors due to degeneracy
with age or stellar metallicity, and due to other possible systematic
uncertainties such as the position matching, and in the assumption of
star formation history and stellar metallicity. The discussion will be
expanded more in Y. Ono et al. (in preparation).

Although the dust extinction of our LAEs is rather high compared with 
those of higher-$z$ LAEs ($E(B-V)=0.00-0.07$ at $z\sim 3$; e.g., 
\citealt{nilsson2007,ono2010a}), 
such an increasing trend of $E(B-V)$ in LAEs down to $z \sim 2$ 
is also seen in other studies \citep{nilsson2011,guaita2011}.
\citet{nilsson2011} performed SED fitting to $z\sim 2.3$ LAEs 
to find that $A_V$ ($E(B-V)$) varies over $0.0-2.5$ ($0.00-0.61$) 
with an average of $A_V=0.6$ ($E(B-V)=0.15$) and $A_V=1.5$ ($E(B-V)=0.37$) 
for old and young population models, respectively.
\citet{guaita2011} obtained $E(B-V)=0.22^{+0.06}_{-0.13}$ 
for stacked $z\sim 2.1$ LAEs assuming constant star formation history.
These findings combined with our result may indicate 
a strong evolution of dust extinction in LAEs from $z>3$ to $z\sim 2$.

\section{Results and Discussion} \label{sec:discussion}

\subsection{Star Formation Rate} \label{ssec:dis_sfr}

\subsubsection{Deriving SFR from the \Ha\ Luminosity} \label{sssec:SFR_derive}

The \Ha\ luminosity is believed to be the most reliable SFR indicator
of galaxies among those based on the rest-frame UV and optical
spectral features. 
Indeed it is proportional to the birth rate of massive stars
as well as being relatively insensitive to dust extinction 
as compared with UV-continuum.
We measure the SFR of the stacked object in the NB209 sub-region
from its \Ha\ luminosity using the relation \citep{kennicutt1998}:
\begin{eqnarray}
{\rm SFR} [M_\odot \ {\rm yr}^{-1}]
= 7.9\times 10^{-42} L({\rm H}\alpha) \ {\rm erg} \ {\rm s}^{-1}.
\label{eq:SFR_Ha}
\end{eqnarray}

Before applying this relation, however, we have to subtract
the contribution from \NII$\lambda\lambda 6584,6548$ lines
from the observed \HaNII\ luminosity.
It is known that the \NII/\Ha\ ratio varies with metallicity;
indeed, this ratio is used as a metallicity indicator of galaxies,
called the $N2$ index \citep[e.g.,][]{PP2004,maiolino2008}.
In Table \ref{tbl:data_stacked}, we derive $L($\Ha$+$\NII$)$ to be
$8.0^{+1.3}_{-1.2}\times 10^{41}$\,erg\,s$^{-1}$.
We use the metallicity estimated in \S \ref{ssec:dis_metallicity}
to infer $L($\NII$)/L($\Ha$)= 5.7^{+1.7}_{-1.3}\times 10^{-2}$ 
\citep{maiolino2008}. The \Ha\ luminosity is then
$8.0^{+1.3}_{-1.2}\times 10^{41} / (1+5.7^{+1.7}_{-1.3}\times 10^{-2})
= 7.6^{+1.2}_{-1.2}\times 10^{41}$\,erg\,s$^{-1}$,
from which we obtain SFR $= 6^{+1}_{-1}\,M_{\odot}$\,yr$^{-1}$.

With $E(B-V)=0.27^{+0.01}_{-0.03}$ combined with \citet{calzetti2000}'s 
extinction law, we obtain the dust-corrected \Ha\ luminosity 
to be $1.7^{+0.3}_{-0.3}\times 10^{42}$\,erg\,s$^{-1}$, 
which is translated into SFR $= 14^{+2}_{-3}\,M_{\odot}$\,yr$^{-1}$.
This SFR is the first unbiased SFR estimate from \Ha\ luminosity 
for typical $z\sim 2$ LAEs of 
$L$(Ly$\alpha)>1\times 10^{42}$ erg\,s$^{-1}$.
Although \citet{finkelstein2011} have derived SFRs 
for two $z\sim 2.3-2.5$ LAEs from \Ha\ spectroscopy, 
both objects have $L$(Ly$\alpha)\sim 2\times 10^{43}$
erg\,s$^{-1}$, which is about one order of magnitude brighter than 
typical Ly$\alpha$ luminosities of LAEs from narrow-band surveys  
(e.g., \citealt{ouchi2008}), due to pre-selection of bright objects 
for NIR spectroscopy.
\citet{hayes2010} have measured $L$(H$\alpha$) for six LAEs 
with $L$(Ly$\alpha)=(0.3-4.5)\times 10^{42}$ erg\,s$^{-1}$
for which \Ha\ emission is detected.
In contrast to these studies, 
our study is based on a large number ($N=105$) of 
purely Ly$\alpha$-selected galaxies.

SED fitting of the stacked object in the NB209 sub-region
gives a stellar mass of $5\times 10^8\,M_{\odot}$.
Therefore, 
the stacked object has the SFR and the stellar mass of the 
same order of magnitudes of those of LAEs studied earlier 
from SED fitting 
($z\sim 2$: \citealt{guaita2011}, %
$z\sim 3$: \citealt{gawiser2006,gawiser2007,lai2008,nilsson2007,ono2010a}).
Some studies have derived SFRs of LAEs from (dust-uncorrected) UV-continuum
(e.g., \citealt{gronwall2007,ouchi2008,ouchi2010,nilsson2009,nilsson2011,
guaita2010})
which are roughly in the range $1-10\,M_{\odot}$\,yr$^{-1}$, 
comparable to our dust-uncorrected SFR.

We also estimate in a similar manner
the SFRs of the (\HaNII)-detected objects, NB387-C-22326, 
NB387-W-04041, and NB387-W-04492 
to be $23^{+2}_{-3}$, $51^{+2}_{-2}$, 
and $13^{+3}_{-3}$ $M_{\odot}$\,yr$^{-1}$, respectively\footnote{%
For NB387-C-22326, we assume $L$(\HaNII) $=L$(\Ha), since its \OII\ is not 
detected and therefore its metallicity is low enough that 
the contribution of \NII\ to $L$(\HaNII) is considered to be negligible. 
For NB387-W-04492, we adopt solar metallicity from the empirical 
\OII/(\HaNII)-metallicity relation (see \S \ref{sssec:determin_Z}) 
to correct for \NII\ emission. 
For NB387-W-04041, the metallicity is estimated to be 
$12+\log({\rm O/H})\sim 8.1$ with Equation (\ref{eq:OII_Ha_NII}).
All SFRs are calculated assuming dust free.
}.
These SFRs are much larger than that of a typical LAE at $z\sim 2$, but
similar to some bright LAEs at similar redshifts 
\citep{nilsson2011, finkelstein2011}.
This is because our NIR narrowband images are relatively shallow,
and only bright, massive LAEs with high SFR are likely to be detected.
Indeed, \citeauthor{nilsson2011}'s  and \citeauthor{finkelstein2011}'s
LAEs are much more massive than a typical LAE
($\gtrsim 10^{10}\,M_{\odot}$)
with some exceptions like HPS256 \citep{finkelstein2011}  
whose mass can be as small as $6\times 10^8\,M_{\odot}$.
Although the (\HaNII)-detected objects lack Spitzer/IRAC data 
and their stellar masses are poorly constrained by SED fitting,
rough estimate is possible from their $K$-band magnitudes.
\citet{daddi2004}, for instance, derived a relation
between $K$ magnitude and mass for $z\sim 2$ BzK galaxies.
The $K$-band total magnitudes of NB387-C-22326, NB387-W-04041, 
and NB387-W-04492 are $23.28$, $23.02$, and $22.39$, which
correspond to a stellar mass of $(2-3)\times 10^{10}$, 
$(2-4)\times 10^{10}$, and $(4-7)\times 10^{10}\,M_{\odot}$, respectively, 
following \citeauthor{daddi2004}'s simple relation.
Although this is just a rough estimate and the relation itself involves
a certain uncertainty ($\sigma (\Delta\log M_*)\sim 0.2$; 
\citealt{daddi2004}), 
the individually (\HaNII)-detected objects are likely to be 
massive LAEs.

\begin{figure}
\epsscale{1.15}
\plotone{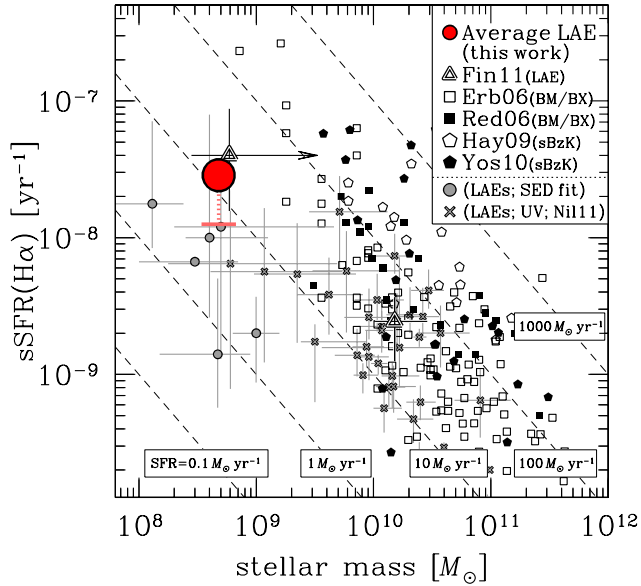}
\caption{
Specific SFR (sSFR) vs. stellar mass
for various types of galaxies at $z\sim 2$.
The red large filled circle indicates our result after correction 
for dust extinction of $E(B-V)=0.27$, while  
the red horizontal bar indicates the sSFR on the assumption of $E(B-V)=0$.
The triangles are LAEs with \Ha\ measurements \citep{finkelstein2011}.
The squares are UV-selected galaxies
(open: \citealt{erb2006b}, filled: \citealt{reddy2006})
and the pentagons are sBzK galaxies
(open: \citealt{hayashi2009}, filled: \citealt{yoshikawa2010}).
All sSFRs are derived from \Ha\ luminosities after dust correction.
A Salpeter IMF is assumed for all objects.
The dashed lines correspond to constant SFRs of 
$0.1, 1, 10, 100, 1000\,M_{\odot}$\,yr$^{-1}$.
The gray filled circles with errorbars are $z=2-3$ stacked LAEs  
whose SFRs are estimated from SED fitting 
(\citealt{guaita2011,nilsson2007,gawiser2006,gawiser2007,lai2008,ono2010a}), 
and the gray crosses are $z=2.3$ LAEs whose SFRs are estimated from 
UV-continuum \citep{nilsson2011}.
\label{fig:SSFRMR_hayashi2009}}
\end{figure}

\subsubsection{Deriving SFR from the \OII\ Luminosity} %
\label{sssec:SFR_derive_OII}

The \OII\ luminosity is also known to be a useful SFR
indicator (e.g., \citealt{gallagher1989, kennicutt1992, kennicutt1998, %
kewley2004, moustakas2006}) and frequently
used to derive SFRs of galaxies at high-$z$, where \Ha\
cannot be accessed from the ground 
(e.g., \citealt{teplitz2003, hopkins2004, takahashi2007}).
The SFR is derived from the \OII\ luminosity using
the relation \citep{kennicutt1998}:
\begin{eqnarray}
{\rm SFR} [M_\odot \ {\rm yr}^{-1}]
= (1.4\pm  0.4)\times 10^{-41} L({\rm [OII]}) \ {\rm erg} \ {\rm s}^{-1}.
\label{eq:SFR_OII}
\end{eqnarray}
In Table \ref{tbl:data_stacked}, we derive $L($\OII$)$ 
in the NB209 sub-region to be $5.3^{+0.4}_{-0.4}\times10^{41}$\,
erg\,s$^{-1}$, which corresponds to SFR $= 7^{+2}_{-2}\,M_{\odot}$\,yr$^{-1}$.
This is consistent with the SFR derived from the \Ha\ luminosity 
when dust free is assumed.
However, if we take into account the dust extinction of
$E(B-V)=0.27^{+0.01}_{-0.03}$, we find that the SFR derived from the 
\OII\ luminosity could be more than twice the SFR derived from 
the \Ha\ luminosity, and they are not consistent within $1\sigma$ errors: 
SFR$_{\rm cor}$(\Ha) = $14^{+2}_{-3}\,M_{\odot}$\,yr$^{-1}$, and 
SFR$_{\rm cor}$(\OII) = $32^{+9}_{-11}\,M_{\odot}$\,yr$^{-1}$. 
Such differences have also been reported for present-day galaxies 
(e.g., \citealt{gilbank2010}) and for high-$z$ galaxies 
(e.g., \citealt{charlot2002, tresse2002}).
The difference can be due to the complicated origin of the \OII\
luminosity. The \OII\ luminosity is not directory proportional to the
ionizing luminosity, and also depends on the chemical abundance
and excitation state of the ionized gas (e.g., \citealt{kennicutt1998, %
kewley2004, moustakas2006}). Indeed, the ratio of
\OII/\Hb\ can be used as a metallicity indicator 
(e.g., \citealt{nagao2006}; see also \S \ref{sssec:determin_Z}).
Alternatively, the uncertainty of the dust correction may cause the
difference of the SFRs. The relatively large dust correction for the
\OII\ luminosity makes it difficult to derive the SFR accurately
(e.g., \citealt{jansen2001}). The difference of the SFRs of our result
may be also due to the overestimates of the dust extinction inferred 
from the SED fitting.
In any case, we use the SFR derived from the \Ha\ luminosity,
which is more directory proportional to the SFR and less affected
by the dust attenuation or other factors (e.g., metallicity), in the
following analysis.

\subsubsection{Mass-sSFR Relation} \label{sssec:MsSFRR}

We plot the stacked object on the specific SFR 
(sSFR $\equiv$ SFR/$M_\star$) vs. stellar mass plane in 
Figure \ref{fig:SSFRMR_hayashi2009}.
Although sBzK and UV-selected galaxies obey a simple
scaling relation between the sSFR and the stellar mass,
our stacked LAE is located below an extrapolation of this 
relation toward lower-mass,
even though the dust extinction of $E(B-V)=0.27$ 
is taken into account in the SFR.
We note that the other LAEs whose SFR and stellar mass are 
derived by SED fitting (gray symbols) are also distributed 
below the extrapolation.
If LAEs are typical of low-mass galaxies,
this figure indicates that low-mass galaxies with 
$M \lesssim 10^{9-10}\,M_{\odot}$ have lower star formation 
efficiencies than extrapolated from more massive galaxies.
This trend, if real, is qualitatively consistent with 
galaxy formation models which predict that less massive 
systems have low star formation efficiencies 
owing to various mechanisms such as feedback from 
supernovae (see, e.g., \citealt{benson2010} for a review).

\subsection{Metallicity} \label{ssec:dis_metallicity}

\begin{figure}
\epsscale{1.15}
\plotone{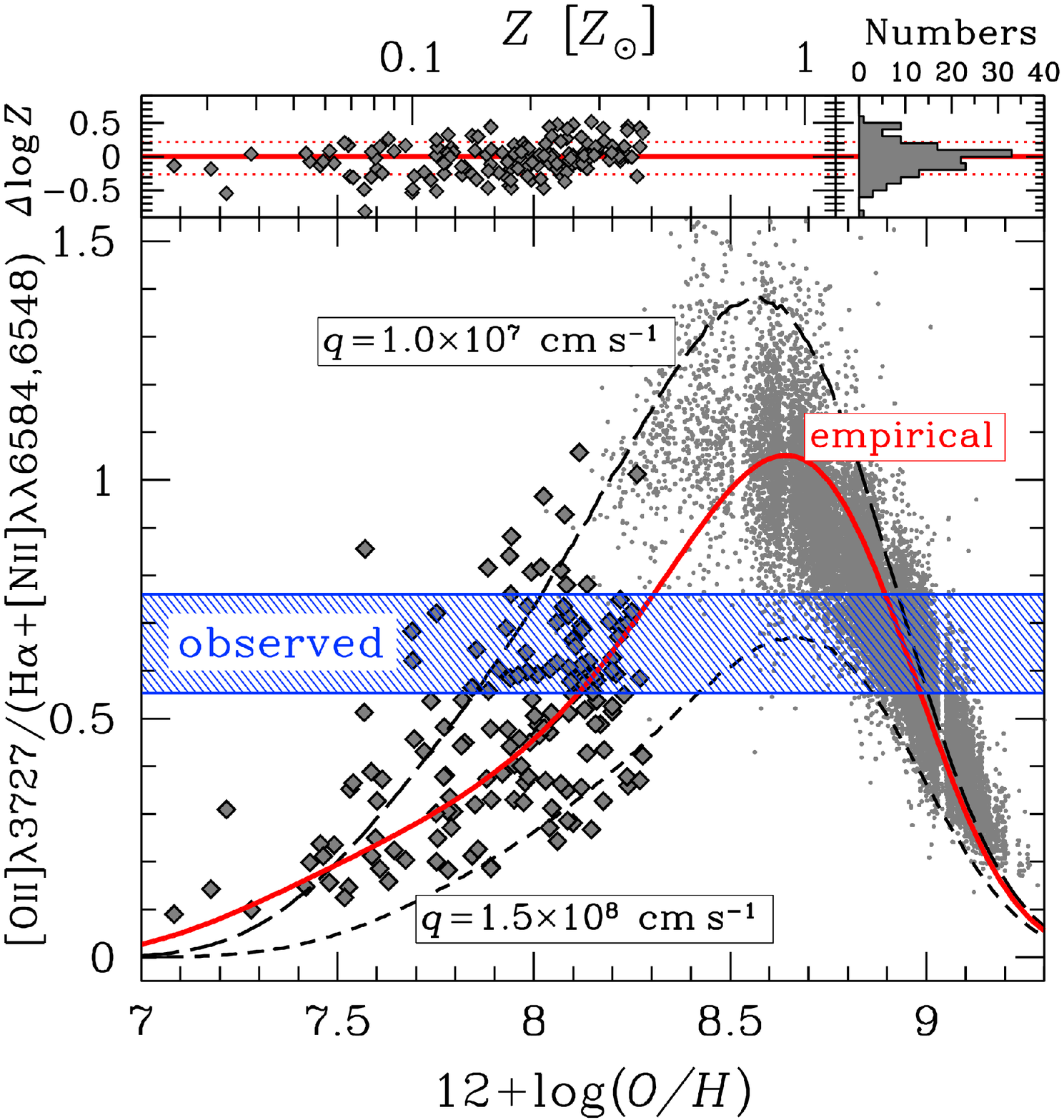}
\caption{
(bottom)
Relation between \OII$\lambda3727$/(\Ha+ \NII$\lambda\lambda6584,6548$)
and gas-phase metallicity.
The red solid line corresponds to the empirical relation given by
Equation (\ref{eq:OII_Ha_NII}) combined with Equation (\ref{eq:Z_function}).
The gray diamonds represent low-metallicity galaxies
compiled by \citet{nagao2006} whose metallicities are inferred through 
the {\it direct $T_e$ method} and then used in \citet{maiolino2008}
to fit the polynomials in Equation (\ref{eq:Z_function})
in the range $12+\log Z<8.3$; 
they are distributed around the best-fit line with a large scatter 
($\Delta \log Z\sim 0.2$).
The gray dots are SDSS galaxies \citep{KD2002} used for the fit 
in the range of $12+\log Z>8.3$ whose metallicities are inferred 
through photoionization models.
The dashed and long-dashed lines correspond to the relations derived
from a combination of stellar population synthesis and photoionization
models with an ionization parameter of
$q=1.5\times 10^8$ and $1.0\times 10^7$\,cm\,s$^{-1}$, respectively
\citep{KD2002}.
We ignore the range $Z>Z_{\odot}$ in this paper,
because LAEs are found to be metal-poor in previous studies
(see \S \ref{sssec:determin_Z}).
The blue shaded area is the range of
\OII$\lambda3727$/(\HaNII$\lambda\lambda6584,6548$)
for the stacked object assuming $E(B-V)=0$.
(top)
Residuals from the best-fit polynomials for the low-metallicity
galaxies (left) and their histogram (right).
The two dotted lines show the rms of the residuals
for $\Delta \log Z>0$ objects (rms$=0.22$)
and $\Delta \log Z<0$ objects ($0.26$), respectively.
\label{fig:Z_OII_Ha_NII} }
\end{figure}

\subsubsection{Constraining the Metallicity} \label{sssec:determin_Z}

An accurate estimate of gas phase metallicity requires knowledge of the 
electron temperature which is provided by comparing
auroral lines to nebular emission lines
(e.g., \OIII $\allowbreak \lambda4363\allowbreak /\lambda5007$; 
{\it direct $T_e$ method}; \citealt{lee2004}).
However, auroral lines are generally weak and it is difficult
to observe them in distant galaxies.
A number of empirical relations between the ratio of nebular lines
and metallicity have been proposed to measure gas-phase metallicities
of distant galaxies.
Among them, we combine the \NII$\lambda6584$/\Ha\ indicator
($N2$ index: e.g., \citealt{PP2004,maiolino2008}) and
the \OII/\Hb\ indicator (e.g., \citealt{nagao2006,maiolino2008})
to derive the metallicity of our object.
Even for the same metallicity index, however, calibrations are often 
different among the authors, leading to different metallicity 
measures as discussed in \citet{nagao2006}.
As for the $N2$ index, \citet{nagao2006}'s calibration 
and \citet{PP2004}'s agree well ($\Delta Z<0.2$\,dex) 
with each other over $7.7<12+\log({\rm O/H})<8.5$, 
while out of this range a large difference is seen 
probably due to the lack of objects for calibration 
in \citet{PP2004} sample.
We use the calibration by \citet{maiolino2008}, 
which is an update of \citet{nagao2006},
based on a low-$Z$ sample ($7.7<12+\log({\rm O/H})<8.3$).

We cannot, however, apply these indicators directly to our object,
because we have only \OII\ and \HaNII\ fluxes.
Instead, we make use the fact that \OII/(\HaNII) ratio can be expressed 
as a combination of the two indicators:
\begin{eqnarray}
\lefteqn{
\frac{{\rm [OII]}}{({\rm H}\alpha+{\rm [NII]\lambda\lambda 6584,6548})}
       } \nonumber \\
 & = & \frac{1}{2.85}\frac{{\rm [OII]}/{\rm H}\beta}
                    {(1+1.33\times{\rm [NII]}\lambda 6584/{\rm H}\alpha)},
\label{eq:OII_Ha_NII}
\end{eqnarray}
where we assume that all lines are dust free
and adopt an intrinsic \Ha/\Hb\ ratio of $2.85$
and an intrinsic \NII$\lambda6584$/$\lambda6548$ ratio of $3.0$
\citep{osterbrock1989}.
The effect of dust extinction will be discussed later.
The metallicity dependence of the two indicators is empirically
approximated by the polynomial:
\begin{eqnarray}
\log R = c_0+c_1x+c_2x^2+c_3x^3+c_4x^4,
\label{eq:Z_function}
\end{eqnarray}
where $R=$ \OII/\Hb\ or \NII/\Ha,
$x$ is the metallicity relative to solar
\citep[i.e., $x=\log(Z/Z_{\odot})=12+\log({\rm O/H})-8.69$,][]{allende2001},
and the coefficients $c_0$ to $c_4$ are taken from Table $4$ of
\citet{maiolino2008}. 
Thus, \OII/(\HaNII) ratio is expressed as a function of metallicity,
as shown in Figure \ref{fig:Z_OII_Ha_NII} by the solid curve.
In this figure, the gray diamonds and gray dots indicate, respectively, 
local low-metallicity galaxies with $12+\log ({\rm O/H}) < 8.3$
\citep{nagao2006} and SDSS galaxies with $12+\log ({\rm O/H}) > 8.3$
\citep{tremonti2004,KD2002} used to calibrate Equation (\ref{eq:Z_function}); 
the metallicities of the former are measured with 
the {\it direct $T_e$ method} \citep{nagao2006}, while those of the 
latter are derived by applying photoionization models to the most 
prominent optical emission lines (\OII, \Hb, \OIII, \Ha, \NII, \SII).
The ratio has a peak at around solar metallicity
and for any given value of \OII/(\HaNII) (except for the peak),
there are two solutions of metallicity, one being subsolar 
and the other being supersolar.
The blue shaded region in Figure \ref{fig:Z_OII_Ha_NII}
corresponds to the observed
\OII/(\HaNII$\lambda\lambda 6584,6548$) ratio
including the $1\sigma$ photometric error: $0.66^{+0.11}_{-0.11}$.
Two metallicity ranges are found to meet the observation:
$12+\log ({\rm O/H}) = 8.21^{+0.10}_{-0.11}$ and $8.94^{+0.05}_{-0.05}$.
The latter range, however, appears to be unlikely,
since it is not consistent with recent spectroscopic observations 
that LAEs have much lower metallicities than the solar value 
\citep{finkelstein2011,cowie2011}. 
Modest dust extinction inferred from SED fitting also 
favors low metallicities
(e.g., \citealt{gawiser2006,gawiser2007,pirzkal2007,lai2008,ono2010a}).
If we rule out the supersolar solution,
the metallicity of our object is estimated to be
$12+\log ({\rm O/H}) = 8.21^{+0.10}_{-0.11}$, or 
$Z/Z_\odot = 0.33^{+0.09}_{-0.07}$.

\begin{deluxetable*}{lrrrrr}
\tablecolumns{6}
\tabletypesize{\scriptsize}
\tablecaption{Coefficients for metallicity indicators
in Equation (\ref{eq:Z_function}) \label{tbl:coefficients_Z}}
\tablewidth{420pt}
\setlength{\tabcolsep}{0.13in}
\tablehead{
\colhead{Flux ratio ($\log R$)} &
\colhead{$c_0$} &
\colhead{$c_1$} &
\colhead{$c_2$} &
\colhead{$c_3$} &
\colhead{$c_4$} 
}
\startdata
$\log$(\NII$\lambda6584$/\Ha)\tablenotemark{(1)} &
$-0.7732$ & $1.2357$ & $-0.2811$ & $-0.7201$ & $-0.3330$ \\
$\log$(\OII$\lambda3727$/\Hb)\tablenotemark{(1)} &
$0.5603$ & $0.0450$ & $-1.8017$ & $-1.8434$ & $-0.6549$ \\ 
$\log$(\OII$\lambda3727$/(\HaNII$\lambda\lambda6584,6548$))\tablenotemark{(2)} & 
$0.0164$ & $-0.1673$ & $-1.9484$ & $-1.8158$ & $-0.6148$ 
\enddata
\tablenotetext{(1)}{%
The coefficients are taken from Table $4$ of \citet{maiolino2008}.
}
\tablenotetext{(2)}{%
The coefficients are determined by a best fit polynomial 
of the combined equation described in Equation (\ref{eq:OII_Ha_NII}).
}
\end{deluxetable*}

\begin{figure}
\epsscale{1.15}
\plotone{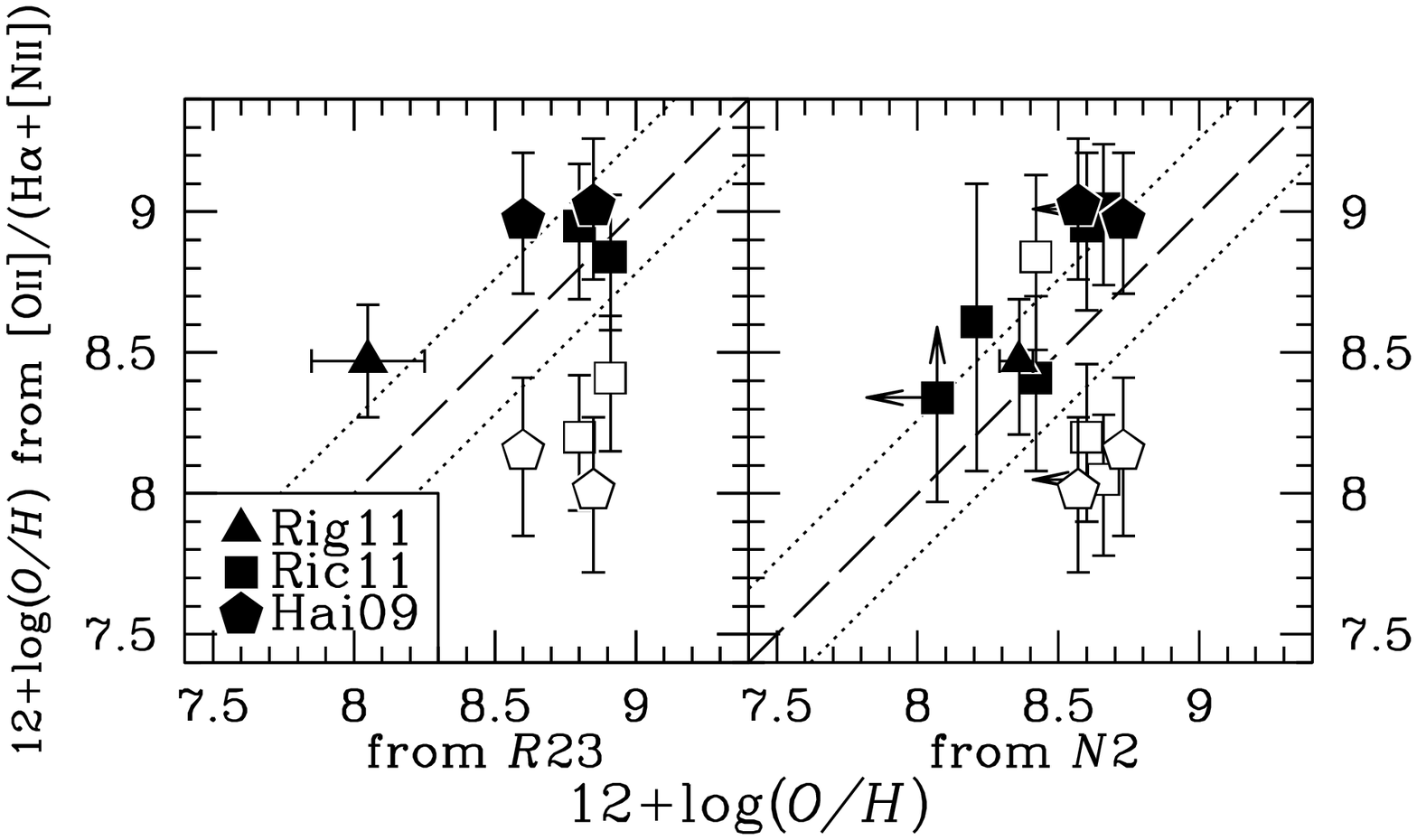}
\caption{
Comparison of the metallicity calculated from our \OII/(\HaNII)
indicator with those from two frequently used indicators,
the $R23$ index (left panel) and the $N2$ index (right panel),
for $z\sim 2$ lensed galaxies taken from the literature
(\citealt{rigby2011} with triangle;
\citealt{richard2011} with squares;
\citealt{hainline2009} with pentagons).
At each panel,
the dashed line is the line of equality,
and the dotted lines show the $1\sigma$ errors associated with
our indicator including
the calibration errors ($\Delta \log Z =+0.22/-0.26$;
see \S \ref{sssec:determin_Z}).
Our indicator has in principle two solutions
(see \S \ref{sssec:determin_Z}), and for galaxies with two solutions
over their $1\sigma$ errors both metallicities are plotted;
solutions with the smaller differences from $Z_{R23}$ or $Z_{N2}$
in black, while those with the larger differences in white.
\label{fig:Z_compare}}
\end{figure}

This metallicity estimate, however,
involves two systematic uncertainties.
First is the uncertainty in the polynomial fit given by
Equation (\ref{eq:Z_function}) for low-metallicity galaxies.
The gray diamonds plotted in Figure \ref{fig:Z_OII_Ha_NII} represent galaxies
with $12+\log ({\rm O/H}) < 8.3$
used in \citet{maiolino2008} to fit the polynomials.
The galaxies are distributed around the solid line (as expected),
but with a large scatter of $\Delta \log Z = +0.22/-0.26$ $(1\sigma)$.
This may suggest that there are large errors in measurements of
line ratios and/or metallicities, since these measurements collected 
from the literature are based on different methods to measure
line ratios \citep{nagao2006}.
Alternatively, such a large scatter may be intrinsic.
Dashed lines in Figure \ref{fig:Z_OII_Ha_NII} represent
relations of metallicity and \OII/(\HaNII) from a combination
of stellar population synthesis and photo-ionization models
with a set of ionization parameter ($q$; \citealt{KD2002}).
From Figure \ref{fig:Z_OII_Ha_NII}, the large scatter could 
be explained by a diversity of ionization parameters
($q\sim (1-10) \times 10^7$\,cm\,s$^{-1}$).
Note that the empirical line can be reproduced by
the \citeauthor{KD2002}'s photo-ionization model
with $q\sim (4-8) \times 10^7$\,cm\,s$^{-1}$.
In any case, we estimate that the calibration error
in our metallicity estimate due to the uncertainty
in the polynomial fit is $\Delta \log Z = +0.22/-0.26$ $(1\sigma)$.
In Figure \ref{fig:Z_compare}, we compare metallicities derived
from our original indicator with those from more commonly used indicators:
the $R23$ index ((\OII$\lambda3727$+\OIII$\lambda\lambda4959,5007$)/\Hb;
e.g., \citealt{tremonti2004} and references therein)
and the $N2$ index to check robustness of the indicator.
We compare the indicators using spectroscopic data of
$z\sim 2$ lensed galaxies \citep{hainline2009,richard2011,rigby2011}.
From the comparison, we see that metallicities from our indicator 
are roughly consistent with those from other indicators within 
$\Delta \log Z = +0.22/-0.26$ $(1\sigma)$.

Second is the uncertainty in dust extinction.
While we have assumed $E(B-V)=0$ to derive Equation (\ref{eq:OII_Ha_NII}),
the SED fit suggests that the stacked object may have $E(B-V)$ up to
$\simeq 0.27$ (see \S \ref{sec:SED_fitting}).
We find that adopting $E(B-V)=0.27$ instead of $E(B-V)=0$ increases
the metallicity estimate by $\Delta \log Z \simeq 0.3$.

When these two systematic errors are taken into account,
the metallicity range of our object is estimated to be
$12 + \log ({\rm O/H}) = 8.21^{+0.10}_{-0.11}\, ({\rm random})\, ^{+0.22}_{-0.26}\, 
(\rm calib)\, = 8.21^{+0.24}_{-0.28}$, 
or $Z/Z_\odot = 0.33^{+0.25}_{-0.16}$, 
assuming $E(B-V)=0$.
If $E(B-V)=0.27$ is adopted, the metallicity increases to
$12 + \log ({\rm O/H}) \sim 8.5^{+0.24}_{-0.28}$, 
or $Z/Z_\odot = 0.7^{+0.5}_{-0.3}$.
Recall, however, that our method logically permits
the possibility of a supersolar metallicity
and that in order to rule out such a possibility,
we will require independent data favoring metal-poor LAEs.
In this sense, the upper limit of the metallicity obtained above
is not as strict as the lower limit.
Considering this, we adopt a conservative conclusion
that the metallicity of the stacked object is no less than
$0.17\,Z_\odot$ $(1\sigma)$, or $0.09\,Z_\odot$ $(2\sigma)$.
This $2\sigma$ lower limit is obtained in the following manner.
The observed \OII/(\HaNII) ratio and its $2\sigma$ negative 
photometric error are $0.66$ and $-0.23$, respectively, thus the 
$2\sigma$ lower limit of the observed \OII/(\HaNII) ratio is $0.43$, 
which corresponds to $12+\log ({\rm O/H})=7.96$.
Therefore, the $2\sigma$ negative statistical error in terms of metallicity 
is $(\Delta \log Z)_{\rm calib} = -(8.21-7.96) = -0.25$.
On the other hand, the $2\sigma$ negative systematic error is 
just twice the $1\sigma$ value $(-0.26)$, 
hence $(\Delta \log Z)_{\rm sys}=-0.52$ $(2\sigma)$.
Therefore, we obtain the $2\sigma$ negative error in terms of metallicity as
$(\Delta \log Z)=-\sqrt{0.25^2+0.52^2} = -0.58$, and the $2\sigma$ 
lower limit of the metallicity as $12+\log ({\rm O/H})=8.21-058=7.63$, 
corresponding to $Z=0.09\,Z_\odot$.
We calculate the $3\sigma$, $4\sigma$, and $5\sigma$ lower limits 
of metallicity in the same manner to obtain 
$12+\log ({\rm O/H})=7.32$, $6.97$, and $6.58$, or $Z=0.04\,Z_\odot$, 
$0.02\,Z_\odot$, and $0.008\,Z_\odot$, respectively.
Although a recent spectroscopic study has placed a weak
upper limit on the metallicity of bright LAEs 
($Z<0.56\,Z_{\odot}$ ($2\sigma$)\footnote{%
We have recalibrated the metallicity given in \citet{finkelstein2011} 
using Equation (\ref{eq:Z_function}). 
The original metallicity upper limit is 
$Z<0.41\,Z_{\odot}$ ($2\sigma$).
}; \citealt{finkelstein2011}),
this is the first lower limit on the metallicity for typical LAEs  
at high-$z$. 
\citet{scannapieco2003} proposed that LAEs may be extremely 
metal poor primordial galaxies, and 
\citet{schaerer2003} also found that large Ly$\alpha$ EWs can be observed 
from extremely metal poor galaxies ($Z/Z_\odot \lesssim 10^{-5}$). 
However, our firm lower limits, e.g., 
$Z>2\times 10^{-2}\,Z_{\odot}$ at the $4\sigma$ level, 
do not support this idea at least for $z=2.2$ LAEs.
For higher redshift ($>3$) LAEs, on the other hand, 
their metallicities may be lower than what we derive 
for $z=2.2$ LAEs, because, as seen in \S \ref{sec:SED_fitting}, 
the amount of dust extinction for LAEs decreases with 
redshift from $z\sim 2$ to $z>3$.

\begin{figure}
\epsscale{1.15}
\plotone{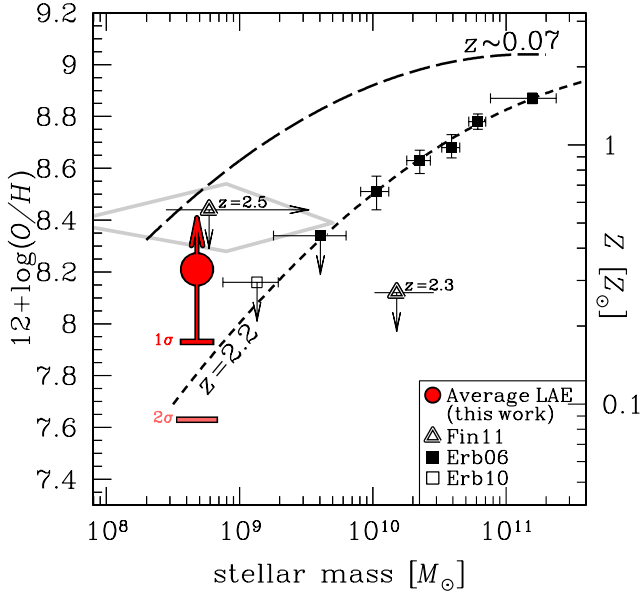}
\caption{
Mass-Metallicity ($M-Z$) relation of UV-selected galaxies
and LAEs at $z\sim 2$.
The red circle with an arrow represents our stacked LAE.
The red ticks show the lower limits of metallicity at 
$1$ and $2\sigma$ levels as labeled next to the ticks.
The open triangles show two LAEs at $z\sim2.3$ and $2.5$ from 
\citet{finkelstein2011}
whose upper limits denote the $2\sigma$ confidence limits.
The curves indicate the $M-Z$ relation observed at
$z\sim 0.07$ \citep[long-dashed;][]{KE2008} and
$z\sim 2.2$  \citep[dashed;][]{erb2006a}
(best-fit functions are determined by \citet{maiolino2008}).
The filled squares are stacked data points of $z\sim 2.2$ UV-selected
galaxies in six bins in stellar mass\citep{erb2006a}, and the
open square indicates BX418 ($z=2.3$; \citealt{erb2010}),
which is considered to be the most metal-poor UV-selected galaxy.
The gray diamond area shows the median stellar mass and metallicity for 
$z=0.195-0.44$ LAEs and their $1\sigma$ ranges \citep{cowie2011}.
All data have been recalibrated to have the same metallicity scale
\citep{maiolino2008} and IMF \citep{salpeter1955}
so that all results can be directly compared.
\label{fig:MZ}}
\end{figure}

\subsubsection{Mass-Metallicity Relation} \label{sssec:MZ_relation}

\begin{figure}
\epsscale{1.15}
\plotone{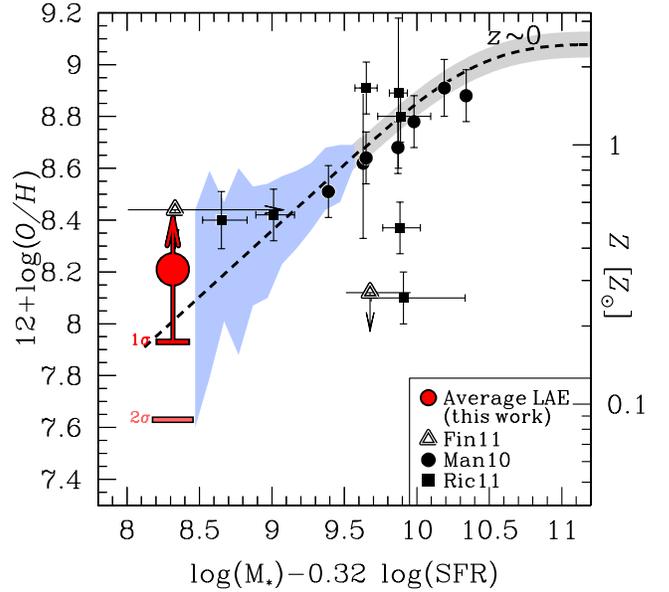}
\caption{
Fundamental $M-Z$ relation proposed by \citet{mannucci2010}.
The dashed curve indicates the best-fit relation
for $z\sim 0$ SDSS galaxies \citep{mannucci2010,mannucci2011},
whose typical distribution ranges are shown
as the gray shaded (for those with
$M_\star\gtrsim 10^{9.2}\,M_{\odot, \rm{chabrier}}$)
and the blue shaded  
($M_\star\lesssim 10^{9.2}\,M_{\odot, \rm{chabrier}}$) 
regions.
The black circles are $z\sim 2.2$ UV-selected galaxies
compiled by \citet{mannucci2010}, the black squares are $1.5<z<2.5$ lensed
galaxies \citep{richard2011}, and the open triangles show two
LAEs at $z\sim2.3$ and $2.5$ from \citet{finkelstein2011}.
The red circle with an arrow is our result; it is located near a smooth
extrapolation of the fundamental $M-Z$ relation toward lower stellar masses.
\label{fig:FMR}}
\end{figure}

Figure \ref{fig:MZ} shows the observed mass-metallicity ($M-Z$) relation
of star-forming galaxies at local and $z\sim 2.2$
compiled by \citet{maiolino2008}:
\citet{KE2008} for $z\sim 0.07$, and 
\citet{erb2006a} for $z\sim 2.2$.
For $z\sim 2.2$, the relation is derived based on stacked galaxies
with $M\gtrsim 3\times 10^9\,M_{\odot}$.
The line below the observed limit of stellar mass is
therefore a smooth extrapolation of the best-fit function.
The $M-Z$ relation is found to evolve with redshift in the sense
that the metallicity at a given stellar mass decreases
with increasing redshift up to $z\sim 3.5$
\citep{maiolino2008,mannucci2009}.

The red circle with an upward arrow represents our stacked object
in the NB209 sub-region which has
a stellar mass of $5\times 10^8\,M_\odot$
and a metallicity of $Z \gtrsim 0.09\,Z_\odot$ ($2\sigma$).
The large circle represents the central value, which is derived assuming 
$E(B-V)=0$, and the two ticks stand for the $1\sigma$ and $2\sigma$ 
lower limits.
Note that our study is the first
to place a lower limit to the metallicity of
$z \sim 2$ galaxies with stellar masses below $10^9 M_\odot$,
well below the lower-mass limit of the previous studies.
The two triangle symbols correspond to the two LAEs at $z \sim 2.3$ and $2.5$
by \citet{finkelstein2011}; for both objects the metallicity
estimate is an upper limit.
The less massive one of the two has a stellar mass
comparable to our object.
If \citeauthor{finkelstein2011} observed a typical LAE,
then the combination of their results with ours suggests that
LAEs with $\lesssim 10^9 M_\odot$ at $z\sim 2$ have metallicities
in the range $0.09 \lesssim Z/Z_\odot \lesssim 0.56$
with the $95$\,\% confidence level.

Comparison of our object with the $M-Z$ relation at $z=2.2$ reveals that
our object is consistent with a smooth extrapolation toward 
lower masses of the $M-Z$ relation at the $1.5\sigma$ level. 
However, we also note that the central value of our object is 
$0.4$ dex larger than the extrapolation. 
This offset might be real, 
since our metallicity estimate is a conservative lower limit.
If real,
there are two possible explanations of this offset.
One is that the slope of the $M-Z$ relation may become shallower
below the stellar-mass limit of UV-selected galaxies
and our object is in fact on the relation,
implying that UV-selected galaxies and LAEs
obey a common $M-Z$ relation.
Although \citet{erb2010} have recently found an unreddened,
low-metallicity, and low-mass UV-selected galaxy (BX418)
at $z=2.3$ (black open square) which seems consistent with 
the conventional $M-Z$ relation at $z\sim 2$ \citep{erb2006a},
it is only one object and
its upper limit for UV-selected galaxies
is still larger than our lower limit.
The other possibility is that
UV-selected galaxies and LAEs obey different $M-Z$ relations
and that the extrapolation is valid only for UV-selected galaxies.
In this case, it is likely that LAEs have relatively high metallicities 
for their masses, unlike the relation estimated for UV-selected galaxies.
Although this appears to be inconsistent with
the {\it conventional} picture that LAEs are the most metal-poor
population, there are, in fact, few studies which compare
metallicities of LAEs and other galaxies statistically
at the same stellar mass.
Thus, we cannot immediately rule out this possibility.

Recently, \citet{mannucci2010} have found that
the observed dispersion in the $M-Z$ relation is correlated with
the SFR in the sense that galaxies with lower SFRs have
higher metallicities, and that star-forming galaxies
at all redshifts below $z\sim 2.2$ obey a common, single 
$M-Z-$SFR relation.
They referred to the relation as the fundamental $M-Z$ relation.
A similar relation between the three quantities is 
reported by \citet{lalalopez2010}, and 
this trend is also realized for high-$z$ lensed galaxies
\citep{richard2011}, 
Indeed, the SFR of our object is lower than that of UV-selected
galaxies with similar mass, and most interestingly,
we find that our object is located near a smooth extrapolation
of the fundamental $M-Z$ relation toward lower stellar masses 
(Figure \ref{fig:FMR}).
In Figure \ref{fig:FMR}, the dashed curve shows the fundamental 
$M-Z$ relation 
defined by SDSS galaxies \citep{mannucci2010,mannucci2011}, 
and the gray and blue shaded areas show 
the typical distribution ranges of galaxies 
with $M_{\star} \gtrsim 10^{9.2}\,M_{\odot,{\rm chabrier}}$ \citep{mannucci2010}
and $M_{\star} \lesssim 10^{9.2}\,M_{\odot,{\rm chabrier}}$ \citep{mannucci2011}, 
respectively.
Therefore, the offset seen in the $M-Z$ relation may be due to
the relatively low SFR of a typical LAE,
and the offset seen in the mass-sSFR relation may be due to 
the relatively high metallicity 
for its mass compared to the value estimated from the $M-Z$ relation of 
UV-selected galaxies.

\subsection{Ly$\alpha$ Escape Fraction} \label{ssec:dis_EF}

We infer the escape fraction of Ly$\alpha$ photons (\fesc)
from the Ly$\alpha$ and \Ha\ luminosities.
The Ly$\alpha$ escape fraction is an important quantity of LAEs
because it can be used to probe distribution of ISM in LAEs.
Since Ly$\alpha$ photons are resonantly scattered
by neutral hydrogen (HI) gas in the ISM,
\fesc\ is strongly dependent on kinematics and distribution
of the ISM as well as the metallicity of the ISM.
For example, \fesc\ will be larger if some ISM is outflowing
(e.g., \citealt{kunth1998,atek2008,DW2010}).
Clumpy distributions of the ISM also makes \fesc\ larger
(e.g., \citealt{neufeld1991,HO2006,finkelstein2008}).

The Ly$\alpha$ escape fraction is calculated by dividing
the observed Ly$\alpha$ luminosity
by the intrinsic Ly$\alpha$ luminosity produced in galaxy
due to star formation. 
To obtain $L_{\rm int}({\rm Ly}\alpha)$,
most studies have used SFRs derived from SED fitting or UV continuum
emission.
However, instantaneous SFRs from SED fitting are model-dependent 
and can have errors as large as two orders of magnitude \citep{ono2010a},
and deriving SFRs from UV continua requires large correction
for dust extinction.
As mentioned before, the SFRs are estimated most reliably from
\Ha\ luminosities.

Recently, \citet{hayes2010} have compared luminosity functions
of LAEs and \Ha\ emitters (HAEs) at $z=2.2$
and found 
the volumetrically averaged escape fraction for star-forming galaxies 
to be $\sim 5$\,\%.
This estimate is based on intrinsically different populations
(LAEs and HAEs) and thus it is not clear whether or not
LAEs typically have such low escape fractions.

We estimate \fesc\ of LAEs at $z\sim 2.2$ by directly comparing
the Ly$\alpha$ luminosity of the stacked object
with its \Ha\ luminosity.
We calculate \fesc\ as:
\begin{eqnarray}
f_{\rm esc}^{{\rm Ly}\alpha}
 \equiv \frac{L_{\rm obs}({\rm Ly}\alpha)}{L_{\rm int}({\rm Ly}\alpha)}
 = \frac{L_{\rm obs}({\rm Ly}\alpha)}
      {8.7 L_{\rm int}({\rm H}\alpha)},
\label{eq:Lya_f_esc}
\end{eqnarray}
where subscripts {\lq}int{\rq} and {\lq}obs{\rq} refer to 
the intrinsic and observed quantities, respectively, 
and we assume Case B recombination
\citep{brocklehurst1971}.

As listed in Table \ref{tbl:data_stacked}, 
$L_{\rm obs}({\rm Ly}\alpha)$ is 
$1.80^{+0.05}_{-0.07}\times 10^{42}$  erg\,s$^{-1}$.
$L_{\rm int}({\rm H}\alpha)$ is derived to be 
$1.73^{+0.28}_{-0.31}\times 10^{42}$\,erg\,s$^{-1}$ 
by correcting $L_{\rm obs}({\rm H}\alpha)$
($7.55^{+1.19}_{-1.15}\times 10^{41}$\,erg\,s$^{-1}$, 
see Table \ref{tbl:data_stacked}) 
for dust extinction of $E(B-V)=0.27^{+0.01}_{-0.03}$.
From Equation (\ref{eq:Lya_f_esc}), 
we find \fesc$=12^{+2}_{-2}$\,\% for the stacked object. 
When dust free is assumed, \fesc\ can be as high as $27^{+4}_{-4}$\,\%, 
which is about six times higher than that derived volumetrically
for star-forming galaxies at $z=2.2$ \citep{hayes2010}. 
These values suggest that for a typical LAE at $z\sim 2$
a relatively large fraction of Ly$\alpha$ photons can
escape the galaxies.
Our value is also much higher than those estimated for
LBGs at $z\sim 3$ \citep[e.g., $\sim 5$\,\% (median);][]{kornei2010},
but similar to that for LAEs at $z=2-3$
($29$\,\% (median): \citealt{blanc2011}, %
$>32$\,\%: \citealt{hayes2010} using LAEs alone, 
$>14$\,\%: \citealt{zheng2011}).

\begin{figure}
\epsscale{1.05}
\plotone{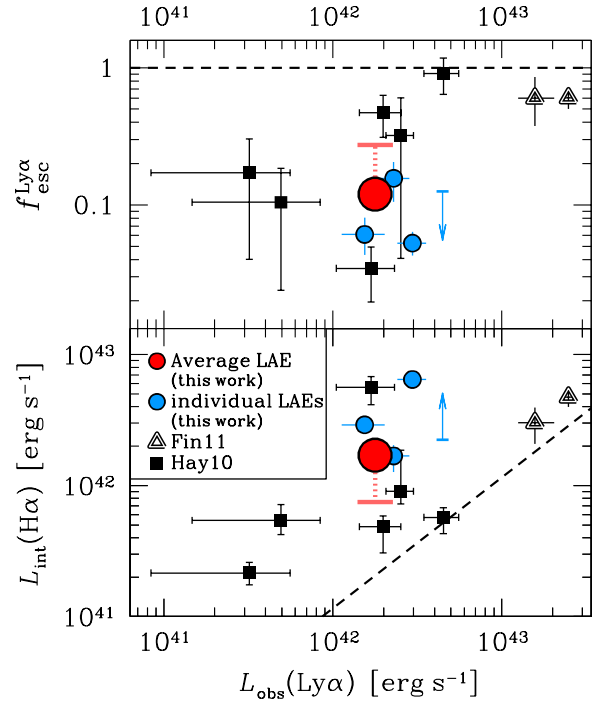}
\caption{
(top) Ly$\alpha$ escape fraction (\fesc) of LAEs against the observed
Ly$\alpha$ luminosity.
The red filled circle indicates our stacked LAE after correction
for dust extinction of $E(B-V)=0.27$,
and the red horizontal bar is the case assuming $E(B-V)=0$.
The cyan circles are three individually \Ha\ detected LAEs;
for them we assume $E(B-V)=0$, implying that their \fesc\
values are upper limits; if they have $E(B-V)=0.27$ as
in the case of the stacked object, \fesc\ decreases by $56$\,\% 
as indicated by the cyan arrow.
The filled squares show $z=2.2$ LAEs which are also selected as
\Ha\ emitters by a double narrowband survey \citep{hayes2010}.
The open triangles denote LAEs at $z\sim 2.3$ and $2.5$ whose Ly$\alpha$
and \Ha\ luminosities are spectroscopically measured \citep{finkelstein2011}.
The dashed line shows \fesc$=1$.
(bottom) Intrinsic \Ha\ luminosity against the observed Ly$\alpha$ luminosity.
The meanings of the symbols and the line are the same as in the top panel.
\label{fig:Lya_f_esc}}
\end{figure}

It is intuitively unreasonable that galaxies with $E(B-V)=0.27$ can escape 
as high as $12$\,\% of Ly$\alpha$ photons.
Indeed, \citet{kornei2010} find that $z\sim 3$ LBGs with $E(B-V)=0.2-0.3$ 
have \fesc\ of just a few percents.
As described in \S \ref{sec:introduction}, however, 
there is a possibility that dust does not always 
attenuate Ly$\alpha$ emission, but rather plays an 
important role in escaping Ly$\alpha$ photons from galaxies. 
In order to quantify the effect of dust on \fesc, we introduce 
the parameter $q$ following \citet{finkelstein2008}, 
which is defined as $q=\tau({\rm Ly}\alpha)/\tau_{1216}$, 
where $\tau({\rm Ly}\alpha)$ and $\tau_{1216}$ are defined as 
$e^{-\tau({\rm Ly}\alpha)} = L_{\rm obs}({\rm Ly}\alpha)/L_{\rm int}({\rm Ly}\alpha)$
and $e^{-\tau_{1216}} = 10^{-0.4k_{1216}E(B-V)}$ 
with $k_{1216}$ being the extinction coefficient at $\lambda=1216$\,\AA.
Small values ($q < 1$) mean that Ly$\alpha$ photons suffer 
less attenuation by dust than UV-continuum photons, as expected 
for a clumpy distribution of the ISM (e.g., \citealt{neufeld1991}) 
or special kinematics of the ISM (e.g., outflows; \citealt{kunth1998}), 
while large values ($q>1$) means that Ly$\alpha$ photons 
are more heavily attenuated by dust.

The $q$ parameter is expressed in terms of \fesc\ and $E(B-V)$ as: 
\begin{eqnarray}
 q = \frac{-\log(f_{\rm esc}^{{\rm Ly}\alpha})}{ 0.4 k_{1216} E(B-V)}. \label{eq:fesc_ebv}
\end{eqnarray}
Using $k_{1216}=11.98$ \citep{calzetti2000} we obtain 
$q=0.7^{+0.1}_{-0.1}$.
Similar results have been obtained from other studies. 
For example, \citet{hayes2010} obtained $q\simeq 1-1.5$ for $z=2.2$ LAEs, 
and \citet{blanc2011} found a median of $q=0.99$ for their 
$z=2-4$ LAE sample.

We note again that the $E(B-V)$ value of our stacked object 
estimated by SED fitting may 
suffer additional errors due to possible systematic uncertainties 
(see \S \ref{sec:SED_fitting}), so may the $q$-value.
However, even when we take an extreme case of $E(B-V)=0.1$,
$q$ increases only up to $\sim 1.5$, 
implying that very large $q$ ($ \gg 1$) are unlikely. 
Our result thus favors models of LAEs' ISM being outflowing, 
or possessing a multi-phase clumpy distribution, or both, 
rather than in a homogeneous, static distribution.

We also apply Equation (\ref{eq:Lya_f_esc}) to the individually detected
objects, NB387-C-22326, NB387-W-04041, and NB387-W-04492 
to obtain \fesc$=6^{+2}_{-2}$, $5^{+1}_{-1}$, and $16^{+5}_{-5}$\,\%, 
respectively.
These values are derived from dust uncorrected \Ha\ luminosities, 
and hence are upper limits.
Figure \ref{fig:Lya_f_esc} shows \fesc\ against
the observed Ly$\alpha$ luminosity
for our LAEs (the red circle for the stacked object and
the cyan circles for the individually detected objects) together with those
LAEs taken from the literature for which \Ha\ emission is
individually detected: 
six LAEs at $z\sim 2.2$ whose \Ha\ luminosities are estimated
from narrowband imaging \citep{hayes2010} and
two LAEs with \Ha\ spectra  at $z\sim 2.3$ and $2.5$ \citep{finkelstein2011}.
It is found that most of the LAEs with individual \Ha\ detection
have \fesc$\gtrsim 10$\,\%,
with objects brighter in Ly$\alpha$ luminosity tending to
have higher escape fractions.
It is interesting that our stacked object is roughly on this trend,
since LAEs with individual \Ha\ detection should be biased
toward higher \Ha\ luminosities.
It may imply that most LAEs (irrespective of \Ha\ detection) have
relatively high \fesc, at least higher than 
the volumetrically averaged star-forming galaxies \citep{hayes2010}.
The trend of increasing \fesc\ with Ly$\alpha$ luminosity
seems to be natural because galaxies with higher \fesc\ have
brighter Ly$\alpha$ luminosities and thus are more easily selected
as LAEs.
Finally, we note that recent studies have shown that a fraction of 
Ly$\alpha$ is emitted from diffuse outer halos of galaxies 
(e.g., \citealt{steidel2011}).
We estimate the strength of Ly$\alpha$ from the color $u^*-$NB387 with 
a certain aperture, therefore the escape fraction we derive is 
only the fraction escaping from the central region of the LAEs.

\section{Conclusions} \label{sec:conclusion}

We have presented the results of the first detection
of \OII\ and \Ha\ emission from a typical Ly$\alpha$ emitter (LAE) 
at $z=2.2$ using a stacking analysis of a sample constructed from
our Subaru/Suprime-Cam narrowband (NB387) survey
in the Subaru/\xmm\ Deep Survey field.
The redshift $z=2.2$ is unique, because \OII\ and \Ha\ lines
fall into NIR wavelengths where OH-airglow is very weak.
We found $919$ LAE candidates in this field.
Follow-up spectroscopy was made for $30$ candidates selected 
to cover wide ranges of NB387 magnitudes and $u^*-$NB387 colors. 
Among the $13$ out of $30$ candidates with NB$387<25$, 
$10$ were confirmed as $z\sim 2.2$ LAEs. 
No emission line was detected for the remaining $20$ with 
NB$387>25$ due to the lack of sensitivity.

The near infrared observation was made to detect \OII\ and \Ha\ 
lines by the NewH$\alpha$ Survey (Lee et al. in preparation) 
with KPNO/NEWFIRM, using the narrowband filters NB118 and NB209, 
respectively;
$561$ LAEs are located in the area covered by NB118
(NB118 sub-sample), among which
$105$ have also NB209 imaging (NB209 sub-sample).

Only seventeen and seven candidates are individually detected 
in NB118 and NB209, respectively.
However, a stacking analysis of a large number of undetected LAEs 
yielded statistically significant detection in both narrowbands. 
We estimated the \OII\ and \HaNII\ EWs and fluxes of the stacked object 
using Monte Carlo simulations, and used the estimates to derive the SFR, 
gas phase metallicity, and Ly$\alpha$ escape fraction (\fesc) of 
a typical LAE. Our main results probed by the triple narrow-band survey 
are summarized as follows.

\begin{itemize}
\item %
The \Ha\ luminosity of the stacked object,
after correction for a contribution from \NII\ lines 
(estimated to be modest) to the \HaNII\ photometry and 
for a dust extinction of $E(B-V)=0.27^{+0.01}_{-0.03}$ 
(derived from SED fitting), 
is $1.7^{+0.3}_{-0.3}\times 10^{42}$\,erg\,s$^{-1}$,
which corresponds to a SFR of $14^{+2}_{-3}\,M_{\odot}$\,yr$^{-1}$.
This is the first estimate of the SFR of a typical LAE at high-$z$
based on the \Ha\ luminosity.
Adopting an stellar mass of $5\times 10^{8}\,M_{\odot}$ derived
from SED fitting,
we plot the stacked object on the specific SFR (sSFR) vs.
$M_\star$ plane,
and find that our stacked object is located below a simple extrapolation
toward lower-masses of the observed sSFR - $M_\star$ relation of
$z\sim 2$ BzK and UV-selected galaxies. 
This trend is also evident for LAEs at similar redshifts 
whose SFRs are inferred by SED fitting.
This indicates that low-mass galaxies with
$M \lesssim 10^{9-10}\,M_{\odot}$ have lower star formation
efficiencies than expected from massive galaxies.

\item %
We use the line ratio \OII/(\HaNII) as a metallicity indicator,
and find that the metallicity of the stacked object is
no less than $0.09\,Z_\odot$ at the $2\sigma$ level.
This is the first constraint on the metallicity of
a typical LAE at high-$z$, and this relatively high lower-limit
does not support, at least at $z\sim 2$,
the hypothesis that LAEs are extremely metal poor 
($Z<2\times 10^{-2}\,Z_{\odot}$) galaxies at the $4\sigma$ level.
We plot the stacked object on the mass-metallicity ($M-Z$) plane,
and find that the stacked object is not consistent with 
a simple extrapolation toward lower masses
of the observed $M-Z$ relation of
$z\sim 2$ UV-selected galaxies.
Instead, our result seems to be consistent with the recently
proposed fundamental $M-Z$ relation \citep{mannucci2010} 
for which the relatively low SFR of the stacked object 
is taken into account.

\item %
From the Ly$\alpha$ and \Ha\ luminosities,
we found that the \fesc\ of the stacked object is $12^{+2}_{-2}$\,\%, and 
can be as high as $27^{+4}_{-4}$\,\%, much larger than those inferred for
volumetrically averaged star forming galaxies at $z=2.2$ and 
Lyman-break galaxies at higher-$z$,
but comparable to those of LAEs at $z=2-3$.
We compiled the \fesc\ data of LAEs with \Ha\ emission from the literature,
and found that most LAEs have relatively high \fesc($\gtrsim 10$\,\%),
and that there are a mild trend that brighter LAEs have higher \fesc.
We also found a low $q$ value for our object, 
$q=0.7^{+0.1}_{-0.1}$. 
All these findings indicate that LAEs have some unique mechanisms to
efficiently emit Ly$\alpha$ photons.

\end{itemize}

\acknowledgments
The NB387 data used in this work are collected at the Subaru 
Telescope, which is operated by the National Astronomical Observatory 
of Japan. We sincerely thank the Subaru Telescope staff for their 
great helps of our Suprime-Cam observations. 
We thank Matthew Hayes for kindly providing us
Ly$\alpha$ and \Ha\ data of his LAE sample as well as useful comments 
on escape fraction of Ly$\alpha$ photons.
We also thank Tohru Nagao and Roberto Maiolino who gave us 
their emission-line flux and metallicity measurements of local galaxies
as well as many fruitful comments on metallicity indicators. 
We thank Chris Simpson for providing the radio source catalog 
for the SXDS.
We thank Lucia Guaita for providing us the average SED of 
$z\sim 2$ BX galaxy.
We are grateful to Steven L. Finkelstein, Maritza Lara-Lopez, Filippo Mannucci, 
Kentaro Motohara, and Zhen-Ya Zheng for their helpful comments and discussions.
We thank the referee for his/her lots of helpful comments 
and suggestions which improved this paper.
This work is based in part on observations made with the Spitzer Space
Telescope, which is operated by the Jet Propulsion Laboratory, California
Institute of Technology under a contract with NASA.
Support for this work was provided by NASA through an award issued by
JPL/Caltech.
This work was supported by World Premier International Research Center 
Initiative (WPI Initiative), MEXT, Japan.

{\it Facilities:} %
\facility{Subaru (Suprime-Cam)}, %
\facility{KPNO:Mayall (NEWFIRM)}, %
\facility{CFHT:MegaPrime (MegaCam)}, %
\facility{UKIRT (WFCAM)}, %
\facility{Magellan:Baade (IMACS)}, %
\facility{Spitzer (IRAC)}



\end{document}